\documentclass[apj, numberedappendix]{emulateapj}
\usepackage{graphicx}
\usepackage{amsmath}
\bibpunct{(}{)}{,}{a}{}{,}

\def\solmass{$M_{\sun}$}

\def\solperyr{$M_{\sun}$ yr$^{-1}$}

\newcommand{\SigGas}{$\Sigma_{\mathrm{gas}}$}
\newcommand{\SigSFR}{$\Sigma_{\mathrm{SFR}}$}
\newcommand{\CSigSFR}{$\langle \Sigma_{\mathrm{SFR}} \rangle$}

\begin{document}

\title{Panchromatic Hubble Andromeda Treasury XVIII. The High-mass Truncation of the Star Cluster Mass Function}

\author{L. Clifton Johnson\altaffilmark{1}, Anil C. Seth\altaffilmark{2}, Julianne J. Dalcanton\altaffilmark{3}, Lori C. Beerman\altaffilmark{3}, Morgan Fouesneau\altaffilmark{4}, Daniel R. Weisz\altaffilmark{5}, Timothy A. Bell\altaffilmark{2}, Andrew E. Dolphin\altaffilmark{6}, Karin Sandstrom\altaffilmark{1}, Benjamin F. Williams\altaffilmark{3}}

\email{lcj@ucsd.edu}
\affil{$^{1}$Center for Astrophysics and Space Sciences, University of California, San Diego, 9500 Gilman Drive, La Jolla, CA 92093, USA}
\affil{$^{2}$Department of Physics and Astronomy, University of Utah, 115 South 1400 East, Salt Lake City, UT 84112, USA}
\affil{$^{3}$Department of Astronomy, University of Washington, Box 351580, Seattle, WA 98195, USA}
\affil{$^{4}$Max-Planck-Institut f\"ur Astronomie, K\"onigstuhl 17, 69117 Heidelberg, Germany}
\affil{$^{5}$Department of Astronomy, University of California, Berkeley, CA 94720, USA}
\affil{$^{6}$Raytheon Company, 1151 East Hermans Road, Tucson, AZ 85756, USA}


\begin{abstract}
We measure the mass function for a sample of 840 young star clusters with ages between 10--300 Myr observed by the Panchromatic Hubble Andromeda Treasury (PHAT) survey in M31.  The data show clear evidence of a high-mass truncation: only 15 clusters more massive than $>10^4$ \solmass\ are observed, compared to $\sim$100 expected for a canonical $M^{-2}$ pure power-law mass function with the same total number of clusters above the catalog completeness limit.  Adopting a Schechter function parameterization, we fit a characteristic truncation mass of $M_c = 8.5^{+2.8}_{-1.8} \times 10^3$ \solmass.  While previous studies have measured cluster mass function truncations, the characteristic truncation mass we measure is the lowest ever reported.  Combining this M31 measurement with previous results, we find that the cluster mass function truncation correlates strongly with the characteristic star formation rate surface density of the host galaxy, where $M_c \propto$ \CSigSFR$^{\sim1.1}$.  We also find evidence that suggests the observed $M_c$--\SigSFR\ relation also applies to globular clusters, linking the two populations via a common formation pathway.  If so, globular cluster mass functions could be useful tools for constraining the star formation properties of their progenitor host galaxies in the early Universe.
\end{abstract}


\section{Introduction} \label{intro}

Star cluster populations are observational tracers of star formation activity in galaxies out to $\sim$100 Mpc distances.  By comparing the properties of star cluster populations to the properties of overall star formation activity, studies of nearby galaxies have established that there is a correlation between the star formation rate (SFR) surface density, \SigSFR, and the fraction of stars that form in long-lived star clusters \citep[e.g.,][]{Adamo15, Johnson16_gamma}.  This correlation demonstrates a close connection between star clusters and their formation environment, where the rate of cluster formation is linked to the total SFR, but also to local galactic properties such as gas surface density and interstellar pressure \citep{Kruijssen12}.  One implication of this result is that star clusters can reveal the characteristics of past star formation episodes long after they have ended.  While cluster destruction through evaporation due to two-body relaxation, tidal shocks, and other processes will erode low-mass star cluster populations over time, globular clusters and other massive clusters provide long-lived records of star formation activity.

The mass function of star clusters is another observable property that we can exploit to study episodes of past star formation.  Numerous studies have characterized the mass function of young star clusters using a power-law distribution ($dN/dM \propto M^{\alpha}$) with an index of $\alpha$=$-2.0\pm0.3$ that holds over a wide range of cluster mass \citep[e.g.,][]{Zhang99, Gieles06a, PZ10, Fall12}.  A power-law mass function slope of $-2$ has the notable property that total cluster mass is distributed equally among logarithmic intervals of cluster mass.  This behavior is consistent with predictions for cluster formation via random sampling from a hierarchical gas distribution, and predictions for clump mass distributions from turbulent fractal clouds \citep[see][and references therein]{Elmegreen08conf}. The observed similarity in shape of the young cluster mass function across a wide range of star-forming environments is often cited as evidence in favor of universal (or ``quasi-universal'') descriptions of cluster formation behavior \citep[e.g.,][]{Fall12}.

There is on-going debate as to whether the high-mass ($>$10$^4$ \solmass) portion of the cluster mass function also follows a power-law distribution, or instead turns over and truncates at some maximum cluster mass.  A pure power-law form is often assumed due to the lack of obvious features in smoothly declining cluster mass distributions and limitations imposed by low number statistics at the high-mass end \citep[e.g.,][]{Chandar10-LMC, Whitmore10}.  However, multiple studies have presented evidence in support of an exponential high-mass truncation through direct mass function fitting \citep[e.g.,][]{Gieles09, Adamo15}, through indirect modeling of the most massive and most luminous clusters \citep[e.g.,][]{Bastian08, Bastian12_MF}, or both \citep{Larsen09}.  Modeling the truncated mass distribution using a \citet{Schechter76} function ($dN/dM \propto M^{\alpha} \exp(-M/M_c)$, where $M_c$ is the characteristic truncation mass), these investigations report mass function truncations with $M_c$ of $\sim$10$^5$ \solmass\ in normal star forming galaxies, and larger values ($\sim$10$^6$ \solmass) for the interacting, starburst Antennae galaxies.

A definitive consensus on the behavior of the high-mass end of the cluster mass function has not yet emerged.  Small sample sizes of massive clusters, relatively small differences between predictions for a pure power-law and a Schechter function, and the indirect nature of some analyses all contribute to the lingering uncertainty.  Nonetheless, measured truncation masses appear to increase systematically with star formation intensity, as observed on galaxy-wide scales \citep{Larsen09}, as well as within individual galaxies \citep{Adamo15}.

We present results from an unparalleled study of the star cluster mass function in the neighboring Local Group galaxy M31, based on data from the \textit{Hubble} Space Telescope (HST) obtained by the Panchromatic \textit{Hubble} Andromeda Treasury survey \citep[PHAT;][]{Dalcanton12}.  High spatial resolution imaging from HST resolves individual member stars in M31's star clusters, and we use these observations to measure cluster ages and masses through color-magnitude diagram (CMD) fitting of the cluster's resolved stars.  This approach provides stronger constraints on young cluster properties than those obtained through multi-band SED fitting, and avoids large uncertainties caused by stochastic variations in the integrated light of low-mass clusters \citep[see e.g.,][]{Fouesneau10, Krumholz15}.

We measure the cluster mass function for a well-characterized sample of 1249 young star clusters drawn from the PHAT cluster catalog \citep{Johnson15_AP}.  Robust cluster identifications and catalog completeness determinations combine to yield a sample of clusters that is well-suited for a mass function investigation.  The catalog's $\sim$10$^3$ \solmass\ 50\% completeness limit for young clusters, combined with our well-characterized completeness function, provides an unprecedented range of masses available for mass function fitting.

The cluster population in M31 allows us to analyze the properties of the cluster mass function in a galaxy that falls at the low-intensity end of the galactic \SigSFR\ spectrum, providing valuable leverage for evaluating possible systematic variations of the high-mass truncation of the cluster mass function.  Previous observations have focused on galaxies with moderate star formation activity \citep[e.g., M51 and M83;][]{Gieles09, Adamo15}, as well as high intensity starburst galaxy mergers \citep[e.g., the Antennae;][]{Zhang99, Whitmore10}.  Our study of M31 extends the range of star formation environments analyzed by an order of magnitude in \SigSFR, providing significant leverage on measuring environmentally-dependent variations of cluster mass function truncations.

We structure the paper as follows.  We begin by introducing the PHAT cluster sample and CMD fitting in Section \ref{data}.  Next, we introduce a probabilistic cluster mass function fitting technique in Section \ref{analysis}, and present results in Section \ref{results}.  We compare our M31 results to Schechter mass function measurements from other young cluster systems and discuss the systematic variation of high-mass truncation masses with \SigSFR\ in Section \ref{discuss_ymc}.  In Section \ref{discuss_gc}, we consider the implications that a $M_c$--\SigSFR\ relation may have on the interpretation of old globular cluster systems.  We summarize our results in Section \ref{summary}.

\section{Data} \label{data}

\begin{figure*}
\centering
\includegraphics[scale=0.4]{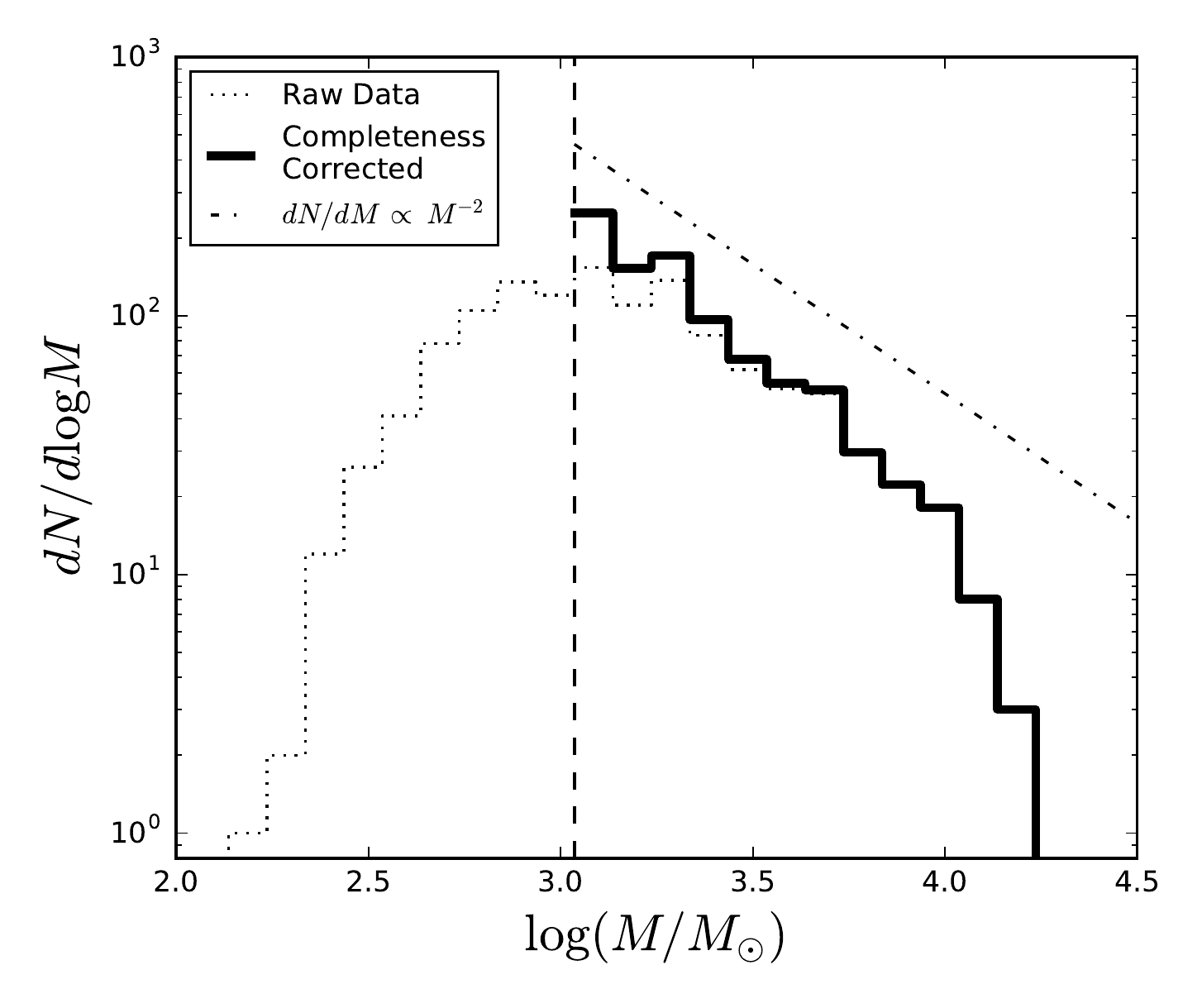}
\includegraphics[scale=0.4]{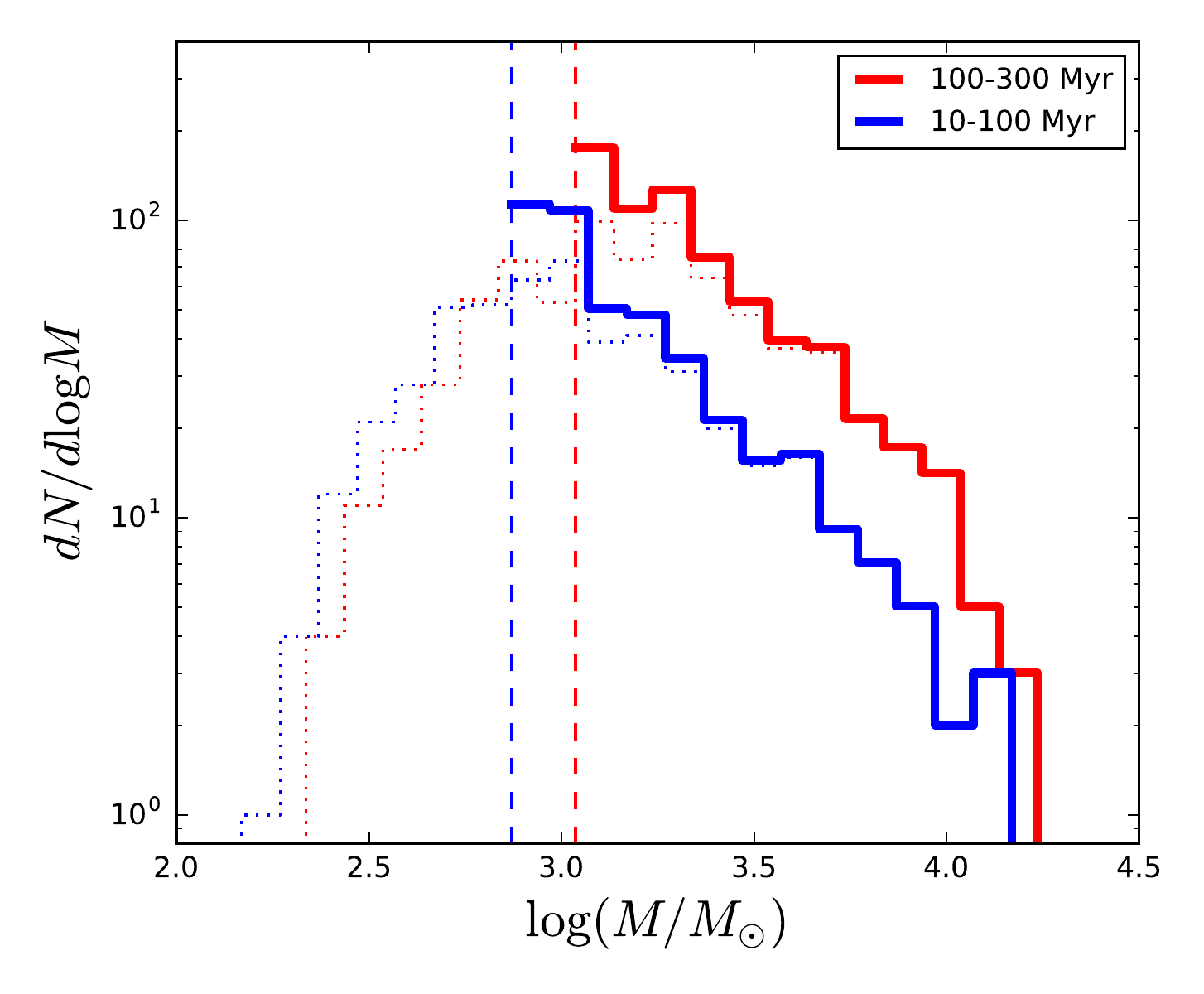}
\includegraphics[scale=0.4]{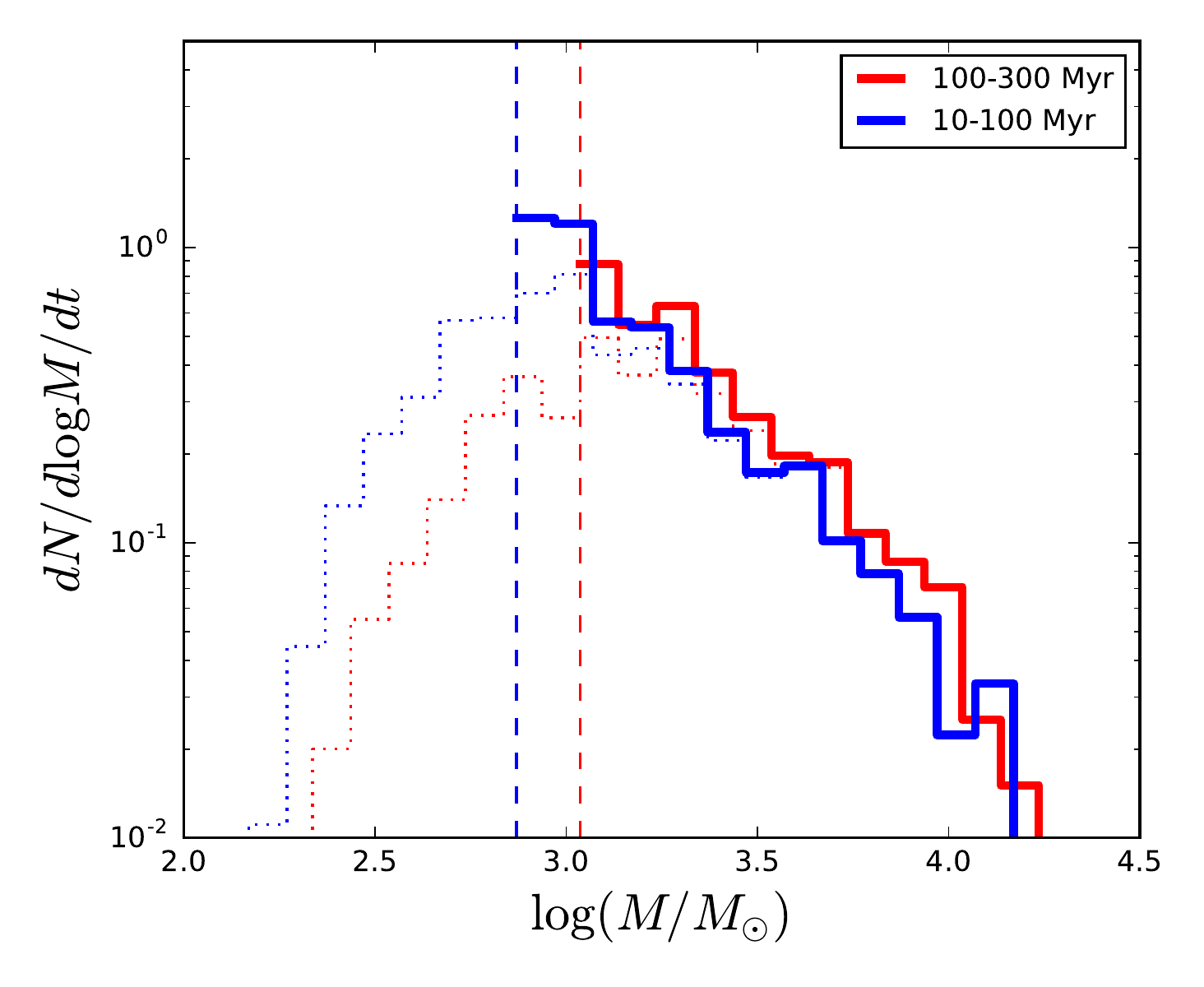}
\caption[PHAT Cluster Mass Function: Observed Distributions]{Observed mass distributions for PHAT star cluster sample, showing raw observed counts (dotted lines) as well as completeness corrected distributions (thick solid lines).  We only consider the portion of the sample that lies above the sample's 50\% completeness limit (vertical dashed lines) for fitting.  The downturn in the raw cluster counts (dotted lines) at low-mass is due to catalog incompleteness. The full 10--300 Myr cluster sample (black) in the left panel shows the same overall shape ($dN/dM \propto M^{-2}$; dash-dotted line) as the younger 10--100 Myr population subset (blue) and the older 100--300 Myr population (red) shown in the center panel. We also plot a duration-normalized version of the age-specific mass functions in the right panel, where the mass distributions are normalized by the width of the age bin to better visualize their similarity.}
\label{fig_data}
\end{figure*}

We draw our cluster sample from the Andromeda Project (AP) cluster catalog \citep{Johnson15_AP}.  This catalog identifies 2753 star clusters that span a wide age and mass range.  The AP catalog was constructed from visual cluster identifications in optical (F475W, F814W; equivalent to $g$ and $I$) images from the PHAT survey data by volunteer citizen scientists, facilitated through a website hosted by the Zooniverse organization.  The final sample of clusters was selected according to a candidate's frequency of identification, where each image was examined by $>$80 AP volunteers.  We adopt a cluster identification threshold that maximizes completeness and minimizes contamination with respect to the expert-derived PHAT Year 1 cluster catalog \citep{Johnson12} and its initial 25\% survey coverage.

The completeness of the cluster catalog was measured using a suite of 3000 synthetic clusters.  Each synthetic cluster was injected into an AP search image and subsequently identified and analyzed in the same way as the genuine clusters; see Section 2.2 in \citet{Johnson15_AP} for detailed properties of the artificial cluster sample.  We compute survey-averaged 50\% completeness limits as a function of cluster mass in two bins in cluster age, 10--100 Myr and 100--300 Myr, following a strategy similar to that used in \citet{Johnson16_gamma} to account for the spatial variation of completeness and star formation across the survey.  First, we bin the synthetic cluster results according to local red giant branch stellar surface density (roughly equivalent to bins of galactocentric radius) to account for the variation of completeness as a function of background stellar density.  Second, we weight the synthetic results in each bin based on local \SigSFR\ to account for the difference in spatial distribution of synthetic clusters versus that of young clusters and star formation.  Third, we calculated a weighted average across the bins of synthetic cluster results, using weights assigned by integrated SFR.  Finally, we fit the weighted and combined completeness results using a logistic function parameterization and find 50\% completeness in mass at 740 \solmass\ for the 10--100 Myr age bin and 1080 \solmass\ for the 100--300 Myr age bin.

Photometric measurements of individual cluster stars were drawn from the catalog of 117 million resolved stars measured as part of the PHAT survey.  The completeness limits of this stellar catalog allow the detection of main sequence stars down to $\sim$3 \solmass.  Please refer to \citet{Dalcanton12} and \citet{Williams14} for full details on the survey's crowded field stellar photometry analysis.  We extract optical (F475W, F814W) CMDs for each cluster, and obtain constraints on cluster parameters through CMD fitting.  We use the MATCH software package to perform maximum-likelihood CMD analysis following techniques described in \citet{Dolphin02}.  For cluster fitting, we adopt a M31 distance modulus of 24.47 \citep[785 kpc;][]{McConnachie05}, a binary fraction of 0.35 with uniform mass ratio distribution, a \citet{Kroupa01} IMF for masses from 0.15 to 120 \solmass, and stellar models from the Padova group \citep{Marigo08} that include updated low-mass asymptotic giant branch tracks \citep{Girardi10}.  We employ a restrictive prior on [M/H] (from $-0.2$ to $0.1$) to constrain solutions to $\sim$$Z_{\sun}$ in an effort to match gas phase metallicity observations within M31 \citep[e.g.,][]{Zurita12}.  Cluster masses from MATCH reflect initial masses, unaffected by mass loss from stellar evolution.  Cluster ages and masses for the PHAT young cluster sample were published as an appendix in \citet{Johnson16_gamma}; we publish a full catalog of cluster parameters, demonstrate the reliability of these results using synthetic cluster tests, and compare CMD-based fits to those derived from integrated light SED fitting in A. Seth et al. (in preparation).

We select a sample of young clusters with ages between 10--300 Myr for mass function analysis.  We adopt a 10 Myr lower limit due to the uncertain and subjective nature of cluster identification at younger ages.  \citet{Gieles11} demonstrate that differentiating between long-lived clusters and rapidly expanding, unbound associations becomes well-defined for ages $>$10 Myr, so we adopt this lower age bound at little expense in terms of integrated star formation and number of clusters.  The upper age bound of 300 Myr is based on the limit where CMD fitting becomes dramatically less precise when the MS turnoff drops below the completeness limit of the stellar photometry.  CMD fitting yields 1249 clusters with best fit ages between 10--300 Myr and masses between 300--20,000 \solmass, where the median age uncertainty is 0.2 dex and the median mass uncertainty is 0.04 dex.  We plot the derived mass distribution for the young cluster sample in the left panel of Figure \ref{fig_data}.

Before we proceed with analysis of the cluster mass function, we note that the age distribution of the PHAT young cluster sample is consistent with a near-constant formation history and little or no cluster destruction \citep{Fouesneau14, Johnson16_gamma}.  The absence of significant cluster mass loss and destruction over the age and mass range we analyze has an important implication: it is safe to assume that the present day mass function we observe can be interpreted as the initial cluster mass function.  In other words, we expect little or no evolution in the shape of the mass function with age due to cluster destruction.  The center and right panels of Figure \ref{fig_data} show that the completeness-corrected mass distributions for age-based subsamples appear qualitatively similar to one another, in agreement with the assumption of no evolution.  When the mass functions for the age subsamples are duration-normalized to account for different bin widths, the two samples also show close agreement in their normalization.  This indicates similar cluster formation rates during these two epochs.  Nonetheless, we will test the assumption of negligible cluster dissolution quantitatively and investigate possible age-dependencies of our results by separately analyzing 10--100 Myr and 100--300 Myr subsamples in addition to the full cluster sample.

\section{Analysis} \label{analysis}

We derive mass function constraints using probabilistic modeling, following an approach similar to that used by \citet{Weisz13} for initial stellar mass function fitting.  The likelihood function of an observed cluster with mass $M$ is given as
\begin{equation}
p_{\rm cluster}(M | \vec{\theta}, \tau) \equiv \frac{1}{Z}\ p_{\rm MF}(M | \vec{\theta})\ p_{\rm obs}(M | \tau),
\end{equation}
where $p_{\rm MF}(M | \vec{\theta})$ is the cluster mass distribution function as defined by the set of parameters $\vec{\theta}$, and $p_{\rm obs}(M | \tau)$ is the observational completeness function, which depends on cluster age, $\tau$.  Finally, $Z$ is the normalization required for $p_{\rm cluster}(M | \vec{\theta}, \tau)$ to properly integrate to 1, given as
\begin{equation}
Z = \int p_{\rm MF}(M | \vec{\theta})\ p_{\rm obs}(M | \tau)\ dM.
\end{equation}

We adopt a \citet{Schechter76} functional form for the cluster mass distribution, whose shape is controlled by two parameters, $\vec{\theta} = \{\alpha, M_c\}$; $\alpha$ is the low-mass power-law index and $M_c$ is the characteristic mass that defines the exponential high-mass truncation.  This distribution follows the form
\begin{equation} \label{eq_s}
p_{\rm MF}(M | \alpha, M_c) \propto M^{\alpha} \exp(-M/M_c).
\end{equation}
Note that the Schechter function simplifies to a simple power-law function ($dN/dM \propto M^{\alpha}$) in the limit that $M_c \to \infty$. We model the age-dependent cluster completeness function using a logistic function, parameterized by the 50\% mass completeness limit, $M_{\rm lim}$, and maximum slope, $a_{\rm lim}$.  The values of the completeness function parameters depend on cluster age, such that ($M_{\rm lim}$, $a_{\rm lim}$)=(740 \solmass, 5.0) for 10--100 Myr old clusters, and ($M_{\rm lim}$, $a_{\rm lim}$)=(1080 \solmass, 5.0) for 100--300 Myr old clusters. To ensure that we are not too sensitive to the completeness corrections, we restrict the model and data to masses greater than the 50\% completeness limit, such that
\begin{equation}
p_{\rm obs}(M | \tau) =
\begin{cases}
\left(1+\exp\left[\frac{-a_{\mathrm{lim}}(\tau)(M-M_{\mathrm{lim}}(\tau))}{M_{\sun}} \right] \right)^{-1}, \\
\hspace{100pt} M > M_{\mathrm{lim}}(\tau)\\
0, \\
\hspace{100pt} \mathrm{otherwise}.
\end{cases}
\end{equation}

We use Bayes' theorem to derive the posterior probability distribution function of the Schechter function parameters, given as
\begin{equation} \label{eq_bayes}
p(\vec{\theta} | \{M_i\}, \tau) \propto p_{\rm cluster}( \{M_i\} | \vec{\theta}, \tau)\ p(\vec{\theta}),
\end{equation}
where $\{M_i\}$ is the set of $N$ cluster masses, $p_{\rm cluster}( \{M_i\} | \vec{\theta}, \tau)$ is the likelihood function for a set of cluster masses, and $p(\vec{\theta})$ is the prior probability for the Schechter function parameters.  The likelihood function for a set of cluster masses is defined as the product of the individual cluster mass probabilities:
\begin{multline} \label{eq_s_like}
p_{\rm cluster}( \{M_i\} | \alpha, M_c, \tau) = \\
\prod_{i=1}^{N} \frac{1}{Z}\ M_i^{\alpha} \exp(-M_i/M_c)\ p_{\rm obs}(M_i | \tau),
\end{multline}
where the normalization term becomes
\begin{equation} \label{eq_s_norm}
Z = \int_{M_{\mathrm{lim}}}^{\infty} M^{\alpha} \exp(-M/M_c)\ p_{\rm obs}(M | \tau)\ dM.
\end{equation}
We adopt uniform top-hat prior probability distributions that generously cover the range of possible parameter values: $-3 \le \alpha \le -1$ and $3 \le \log (M_c$/$M_\sun$) $\le 8$.  These uninformative priors on $\alpha$ and $M_c$ are sufficiently broad to enclose all points in parameter space with non-trivial likelihoods, such that the fitting results are not sensitive to their specific limits.  Finally, we integrate the normalization term numerically during the course of fitting. 

The probabilistic framework we use here for cluster mass function fitting assumes negligible uncertainties on individual cluster masses.  \citet{Weisz13} demonstrate that this simplifying assumption does not significantly bias fitting results in the limit of small (0.1) fractional mass uncertainties.  As the fractional error on the masses increases to 0.5 and beyond, fitting results become more and more affected.  The PHAT CMD-based cluster masses have a median uncertainty of 0.04 dex, and these mass uncertainties are smallest at the high-mass end of the cluster sample where individual masses have the greatest leverage over $M_c$ results.  Therefore, we are confident that the assumption of negligible mass errors does not significantly impact the results presented here. 

\subsection{Power-law Functional Form} \label{powerlaw}

We also adapt this probabilistic framework to fit a non-truncated, power-law functional form of the cluster mass distribution.  For this purpose, we adopt
\begin{equation} \label{eq_pl}
p_{\rm MF}(M | \vec{\theta}) \propto M^{\alpha},
\end{equation}
and power-law equivalents of the likelihood function for the set of cluster masses and its normalization (Eqs. \ref{eq_s_like} and \ref{eq_s_norm}) are given as
\begin{equation} \label{eq_pl_like}
p_{\rm cluster}( \{M_i\} | \alpha, \tau) = \prod_{i=1}^{N} \frac{1}{Z}\ M_i^{\alpha}\ p_{\rm obs}(M_i | \tau)
\end{equation}
and
\begin{equation} \label{eq_pl_norm}
Z = \int_{M_{\mathrm{lim}}}^{\infty} M^{\alpha}\ p_{\rm obs}(M | \tau)\ dM.
\end{equation}

\subsection{Sampling the Posterior Probability Distributions} \label{powerlaw}

We use a Markov Chain Monte Carlo (MCMC) technique to sample the posterior probability distributions of the Schechter and power-law mass function parameters.  In particular, we use the \texttt{emcee}\footnote{\url{http://dan.iel.fm/emcee/}} Python package \citep{ForemanMackey13} and its implementation of an affine invariant ensemble sampler from \citet{GoodmanWeare10}.  For the MCMC calculation, we use 500 walkers, each producing 600 step chains, of which we discard the first 100 burn-in steps.  We report the median value of the marginalized posterior probability distribution function (PDF) for each of the Schechter function parameters, $p(M_c | \{M_{i}\}, \tau)$ and $p(\alpha | \{M_{i}\}, \tau)$, accompanied by a 1$\sigma$ confidence interval defined by the 16th to 84th percentile range of the marginalized posterior.  For the power-law function, we report the median and 1$\sigma$ confidence interval for the single parameter, $\alpha$.

\section{Results} \label{results}

\subsection{Schechter Fitting Results} \label{results_sch}

\begin{figure*}
\centering
\includegraphics[scale=0.7]{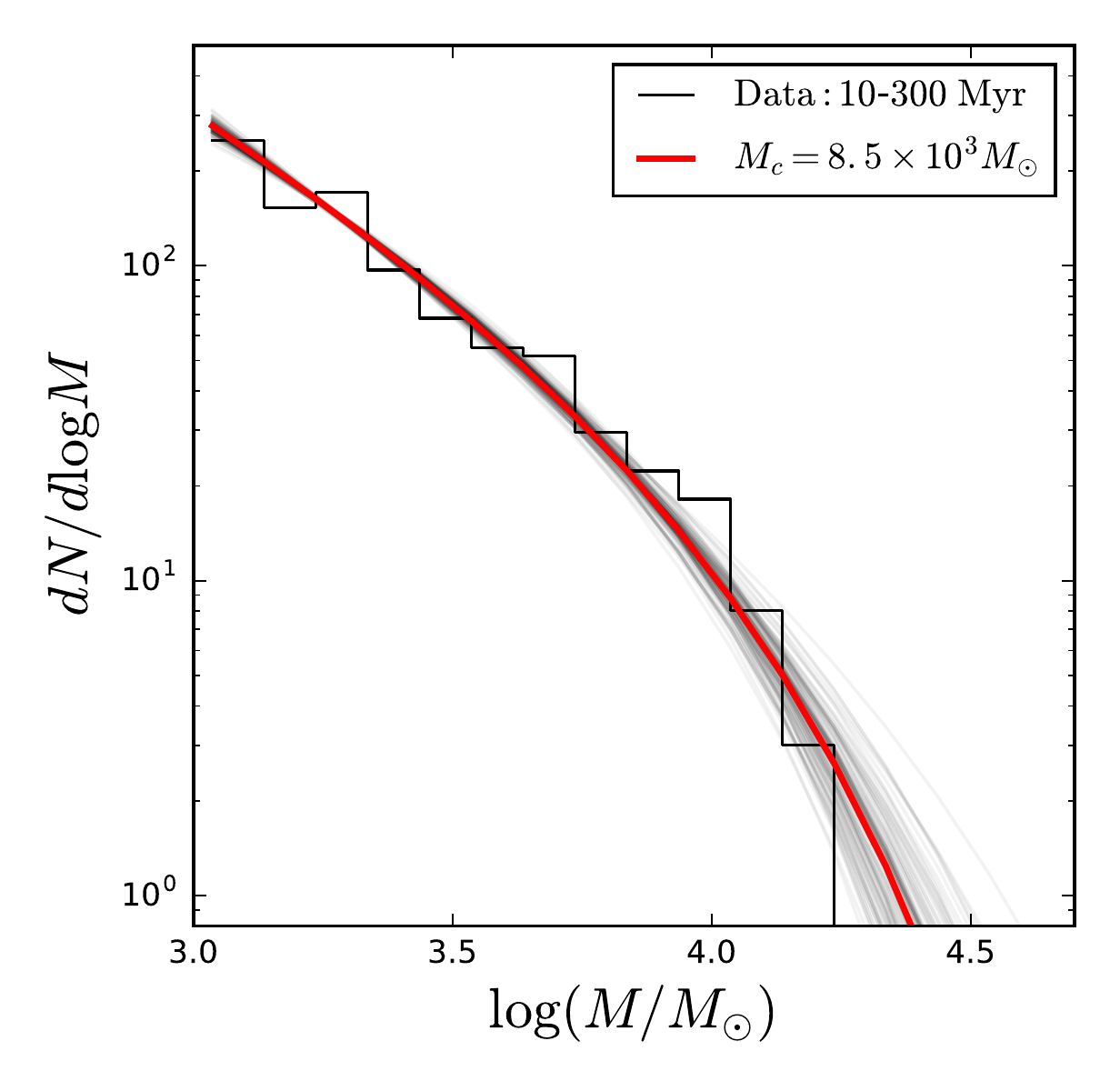}
\includegraphics[scale=0.5]{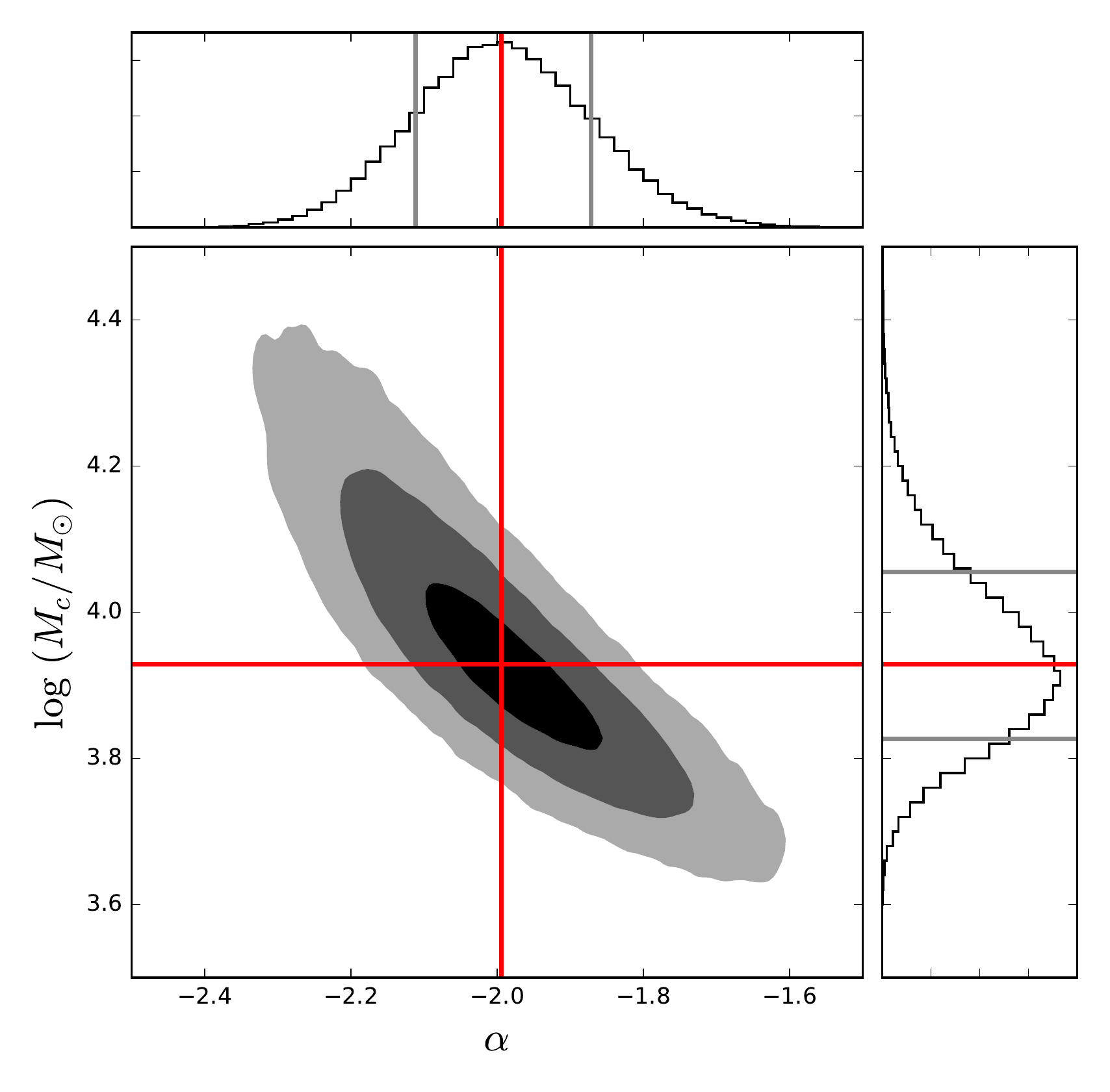}
\caption[PHAT Cluster Mass Function: Schechter Function Fit]{Schechter function fitting results for PHAT cluster sample.  Left: Histogram shows the observed, completeness-corrected cluster mass distribution (black).  We use a binned histogram for visualization purposes only, and the fitting is performed to an unbinned mass distribution.  We plot 100 samples from the posterior PDF to show the variance in Schechter fits (gray) around the well-fit median-selected model (red). Right: Two-dimensional posterior constraints on $\alpha$ and $M_c$, where contours represent 1, 2, and 3$\sigma$ confidence intervals and additional panels show marginalized one-dimensional PDFs for $\alpha$ and $M_c$.}
\label{fig_fit_sch}
\end{figure*}

Schechter function fitting results for the 10--300 Myr PHAT young cluster sample are shown in Figure \ref{fig_fit_sch}, derived for 840 clusters whose masses are greater than the 50\% mass completeness limit of the cluster's age bin.  In the left panel, we compare the observed, completeness-corrected cluster mass distribution to Schechter function fits.  We draw pairs of $M_c$ and $\alpha$ parameter values from the posterior PDF and normalize these functions to match the completeness-corrected number of clusters above the most restrictive completeness limit (from the 100--300 Myr age bin) at 1080 \solmass.  We stress that the binned mass distribution shown here is only used for visualization purposes, and that our results are based on probabilistic fitting of individual, unbinned cluster masses.

We find that the PHAT young cluster sample is well-described by a Schechter function with $M_c$ = $8.5^{+2.8}_{-1.8} \times 10^3$ \solmass\ ($\log M_c$/\solmass = $3.93^{+0.13}_{-0.10}$) and $\alpha$ = $-1.99 \pm 0.12$.  These results are based on the one-dimensional marginalized posterior PDFs, which we present in the right panel of Figure \ref{fig_fit_sch} along with the two-dimensional posterior PDF that shows the covariance between the Schechter function parameters.  The characteristic truncation mass reported here is the lowest value ever obtained for a star cluster population, which is more than an order of magnitude below the $2 \times 10^5$ \solmass\ value derived for a sample of star forming galaxies by \citet{Larsen09}.  The index of the low-mass slope agrees perfectly with the canonical value of $-2$, supporting the notion that the mass function for the M31 PHAT cluster sample is otherwise rather typical at lower cluster mass.

\subsubsection{Testing for Age Dependence} \label{results_age}

\begin{figure}
\centering
\includegraphics[scale=0.72]{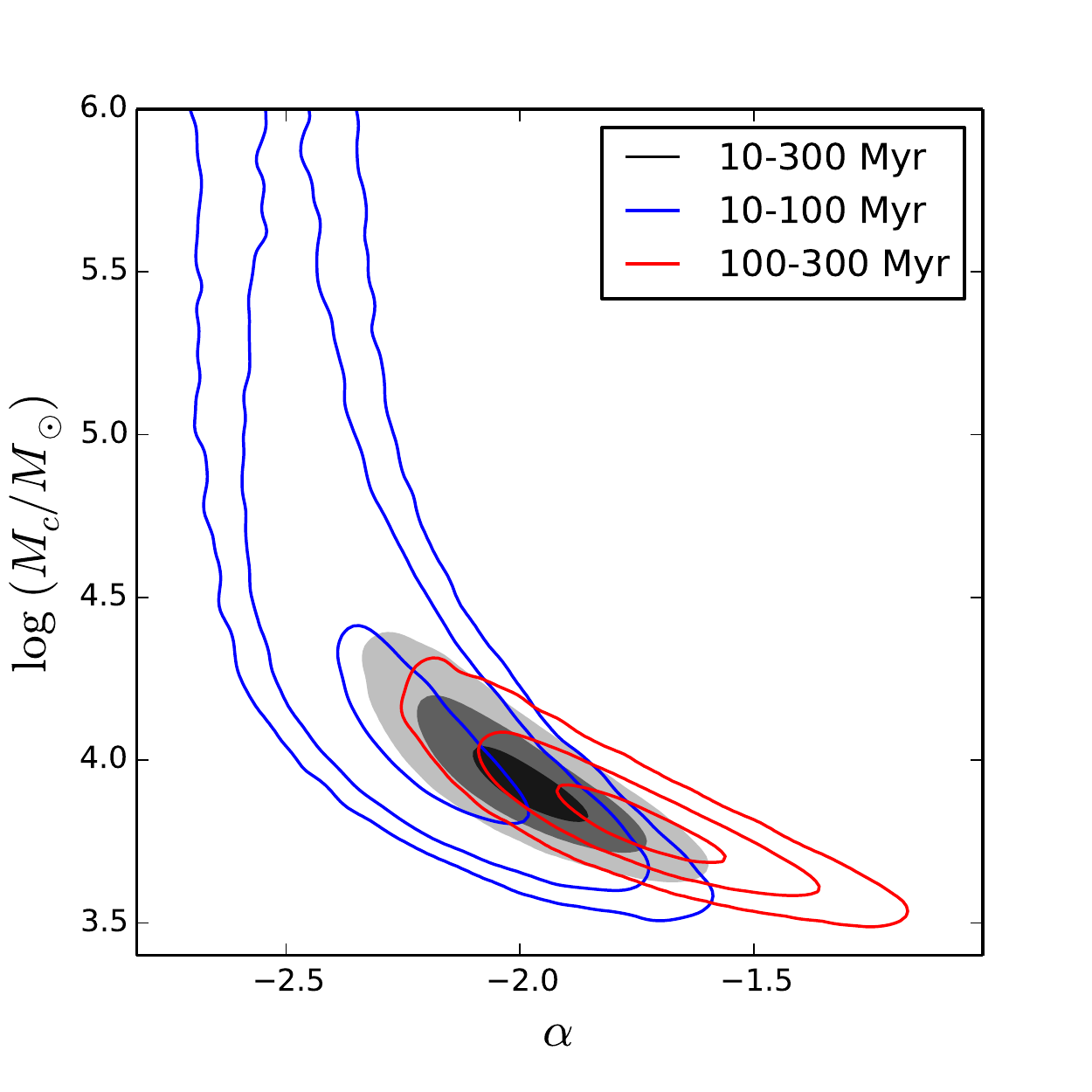}
\caption[$M_c$ Fit: Age Comparison]{Two-dimensional posterior constraints on $\alpha$ and $M_c$ for the 10--100 Myr (blue) and 100--300 Myr (red) cluster samples, overlaid on sample-wide (10--300 Myr) constraints (grayscale).  Contours represent 1, 2, and 3$\sigma$ confidence intervals.}
\label{fig_agecomp}
\end{figure}

A notable signature of mass dependent cluster dissolution is a flattening of the low-mass slope of the cluster mass function with increasing age \citep{Gieles09}.  We test for age-dependence in our Schechter mass function fits by dividing the sample into two age bins: 10--100 Myr and 100--300 Myr.  A comparison of the two-dimensional posterior PDFs for all three cases of age binning is presented in Figure \ref{fig_agecomp}.  Note that the $M_c$ constraint from the younger age bin alone is significantly weaker due to the reduced number of clusters; only 324 clusters in the 10--100 Myr age range lie above the bin's 50\% mass completeness limit.  This demonstrates that our large sample of clusters, obtained by integrating over a wide age range and down to low cluster mass, was key to obtaining a robust result.  Nevertheless, we obtain very similar results for the two separate age bins as we did for the total 10--300 Myr sample, and find no significant age dependence of the mass function shape.

The $\alpha$ constraints for the two age bins show a marginal trend of a flatter slope for older ages, but both bins are also consistent with a single $-2$ power-law slope at $\sim$1.5$\sigma$ confidence.  Therefore, the Schechter function fitting results show no significant or definitive signature of cluster dissolution on $\sim$100 Myr timescales, in agreement with previous PHAT analysis of age and mass distributions \citep{Fouesneau14}.  Together, these results suggest that characteristic cluster dissolution timescales longer than the age range examined here ($>$300 Myr).  We will pursue constraints on the timescales and mass dependence of cluster dissolution in future work (M. Fouesneau et al., in preparation).

\subsubsection{Comparison to Previous Work} \label{results_prev}

Previous studies of the young cluster mass function in M31 did not detect a truncation mass of $\sim$10$^4$ \solmass.  For example, \citet{Vansevicius09} compare their ground-based M31 cluster sample \citep[and the sample from][]{Caldwell09} with a Schechter function distribution and argue that their results are consistent with the \citet{Larsen09} spiral galaxy sample average $M_c$ value of $2\times10^5$ \solmass, although they did not perform any fitting.

There are a number of points to consider when comparing our PHAT results to the work of \citet{Vansevicius09} and \citet{Caldwell09}.  First, these two studies were both significantly limited by the low-mass completeness cutoffs of their catalogs.  \citet{Vansevicius09} and \citet{Caldwell09} have 50\% completeness limits at $\log(M/M_{\sun})$ of 3.7 and 4.0, respectively, which is comparable to the $M_c$ value we measured for PHAT.  Without a full accounting of the cluster population at masses below the knee of the distribution, it is difficult to properly constrain the characteristic truncation mass.  Second, the \citet{Vansevicius09} sample only contains a single $10^5$ \solmass\ cluster at masses greater than $5\times10^4$ \solmass, revealing extremely sparse sampling near their preferred $M_c$ value of $2\times10^5$ \solmass.  Third, both of these works analyze clusters from a broader age range, including clusters with ages between 1--3 Gyr.  We prefer to restrict our analysis to an age regime where we can obtain robust cluster fits from CMD fitting. Fourth, both \citet{Vansevicius09} and \citet{Caldwell09} derive masses using conversions based on fully-sampled mass functions.  As mentioned in the introduction, this strategy can lead to significant mass discrepancies due to the stochastic contribution of luminous evolved members.

Finally, the potential exists that the cluster population surveyed by these previous works, which include clusters that lie on the southwest side of the M31 disk opposite that of the PHAT survey region, might truly represent a different star formation environment with higher intrinsic values of $M_c$.  The southwest portion of M31 hosts the star forming complex NGC206 \citep{Hunter96} and vigorous star formation near the split in the 10 kpc star forming ring \citep{Gordon06}, and is known to host a number of notable massive ($10^4$--$10^5$ \solmass) young clusters \citep[e.g., VdB01;][]{Perina09}.  Indeed, \citet{Elmegreen97} point out that the southwestern portion of the M31 disk hosts a spiral arm segment \citep[S4; also OB79--82 in the parlance of][]{vdB64} with particularly high intensity star formation, highlighting this same region of interest.  With this in mind, we note that our results apply only to the PHAT survey region covering the NE quadrant of M31, and that variations across the disk of M31 are possible.  Further study of the active southwest portion of the M31 disk could provide an interesting counterpoint to the more moderate star formation surveyed by PHAT.

\subsubsection{Fitting of Radially-selected Cluster Subsamples} \label{results_radial}

\citet{Adamo15} present Schechter function fitting results for M83 that show a radial trend in truncation mass, such that $M_c$ decreases with increasing galactocentric radius.  These results motivate us to ask: beyond the survey-wide results presented, can  radial trends in $M_c$ be detected in M31?  Adopting region definitions from \citet{Johnson16_gamma}, we assemble inner disk, 10 kpc ring, and outer disk spatial subsamples.  Unfortunately, the present M31 cluster dataset from the PHAT survey does not provide strong constraints on radial trends in $M_c$ due to low number statistics in regions outside the 10 kpc star-forming region, which dominates the PHAT cluster sample ($>$60\% of the total).  There are only 144 and 82 clusters with masses greater than the 50\% completeness limit in the inner disk and outer disk regions, respectively.  These cluster counts are far smaller than the 324 young cluster sample that yielded weak constraints on $M_c$.  Preliminary analysis yields weak constraints for the outer disk region ($>$0.5 dex uncertainty on $M_c$), and only a lower limit for the inner disk region.  Further analysis is required to confirm the robustness of these fitting results in the low number statistics regime.

\subsection{Power-law Fitting Results and Comparison to Schechter Function} \label{results_pl}

\begin{figure*}
\centering
\includegraphics[scale=0.6]{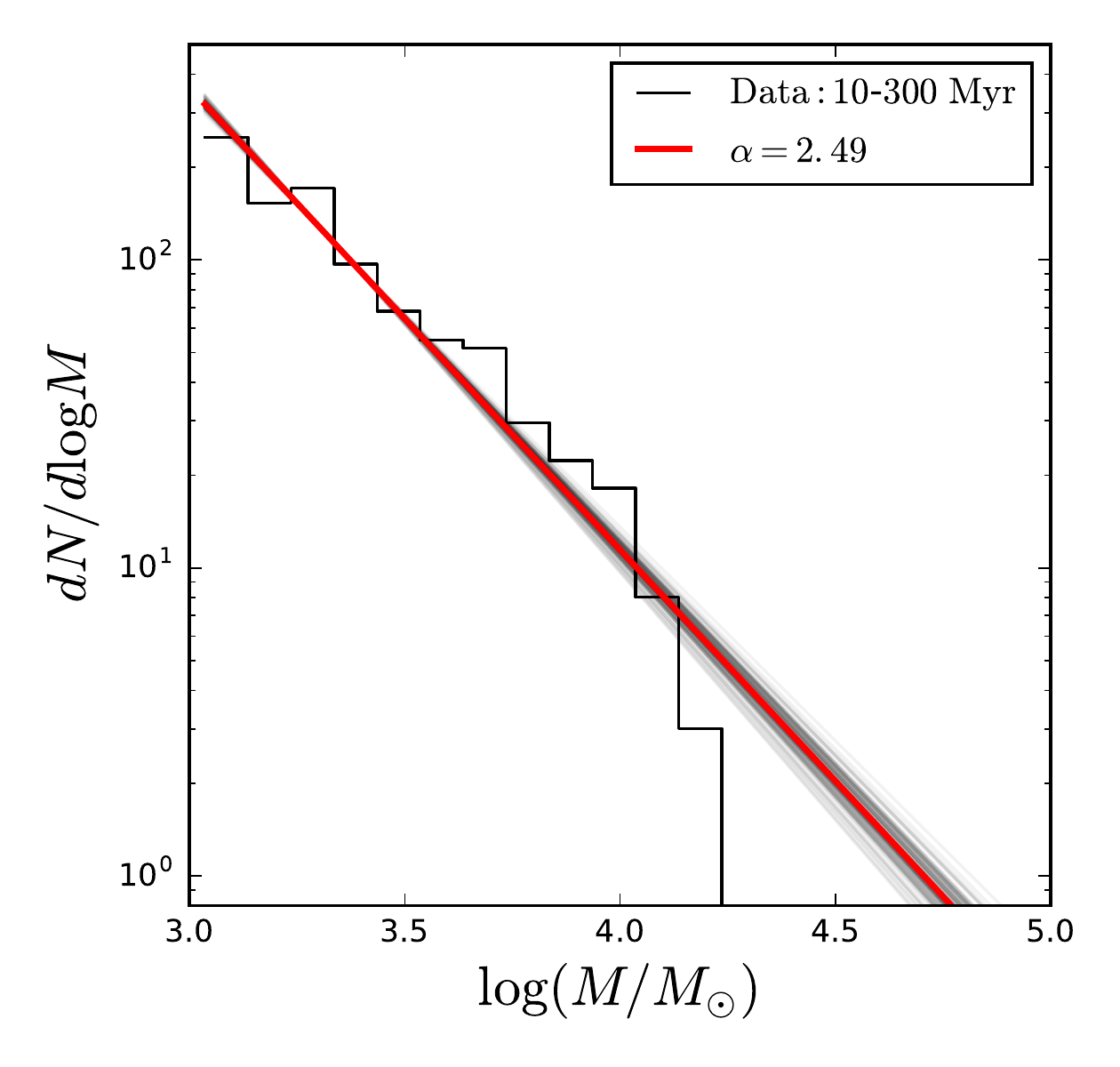}
\includegraphics[scale=0.6]{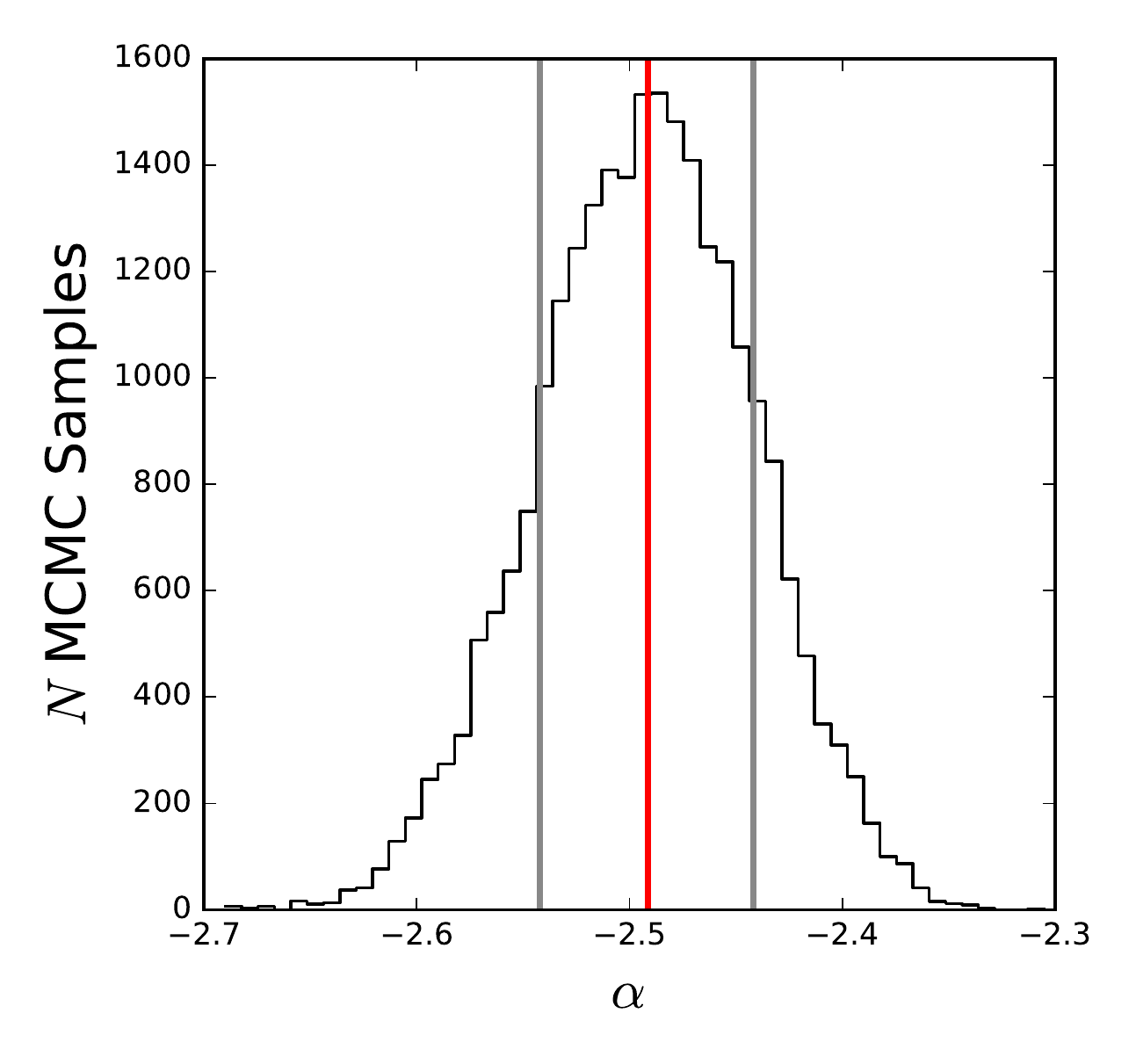}
\caption[PHAT Cluster Mass Function: Power-law Function Fit]{Power-law function fitting results for PHAT cluster sample.  Left: Histogram shows the observed, completeness-corrected cluster mass distribution (black).  We use a binned histogram for visualization purposes only, and the fitting is performed to an unbinned mass distribution.  We plot 100 samples from the posterior PDF to show the variance in power-law fits (gray) around the well-fit median-selected model (red). Right: Posterior PDF for $\alpha$, with gray vertical lines denoting 1$\sigma$ range around the median value (red vertical line).}
\label{fig_fit_pl}
\end{figure*}

While we find that the observed cluster mass distribution is well-described by a Schechter function, we also fit a power-law functional form for comparison. We present power-law fitting results for the PHAT young cluster sample in Figure \ref{fig_fit_pl}.  Similar to Figure \ref{fig_fit_sch}, we compare realizations of the power-law function to the observed mass distribution in the left panel, where we draw $\alpha$ values from the posterior PDF and normalize the functions to match the completeness-corrected number of clusters above a limiting mass of 1080 \solmass.  We find that the PHAT young cluster sample is best described by a power-law function with $\alpha$ = $-2.49 \pm 0.05$, and we plot the posterior PDF for $\alpha$ in the right panel of Figure \ref{fig_fit_pl}.  This fitted power-law index is much steeper than the canonical $-2$ value, and tends to over-predict the number of clusters at masses greater than 10$^4$ \solmass.

\begin{figure*}
\centering
\includegraphics[width=.34\textwidth]{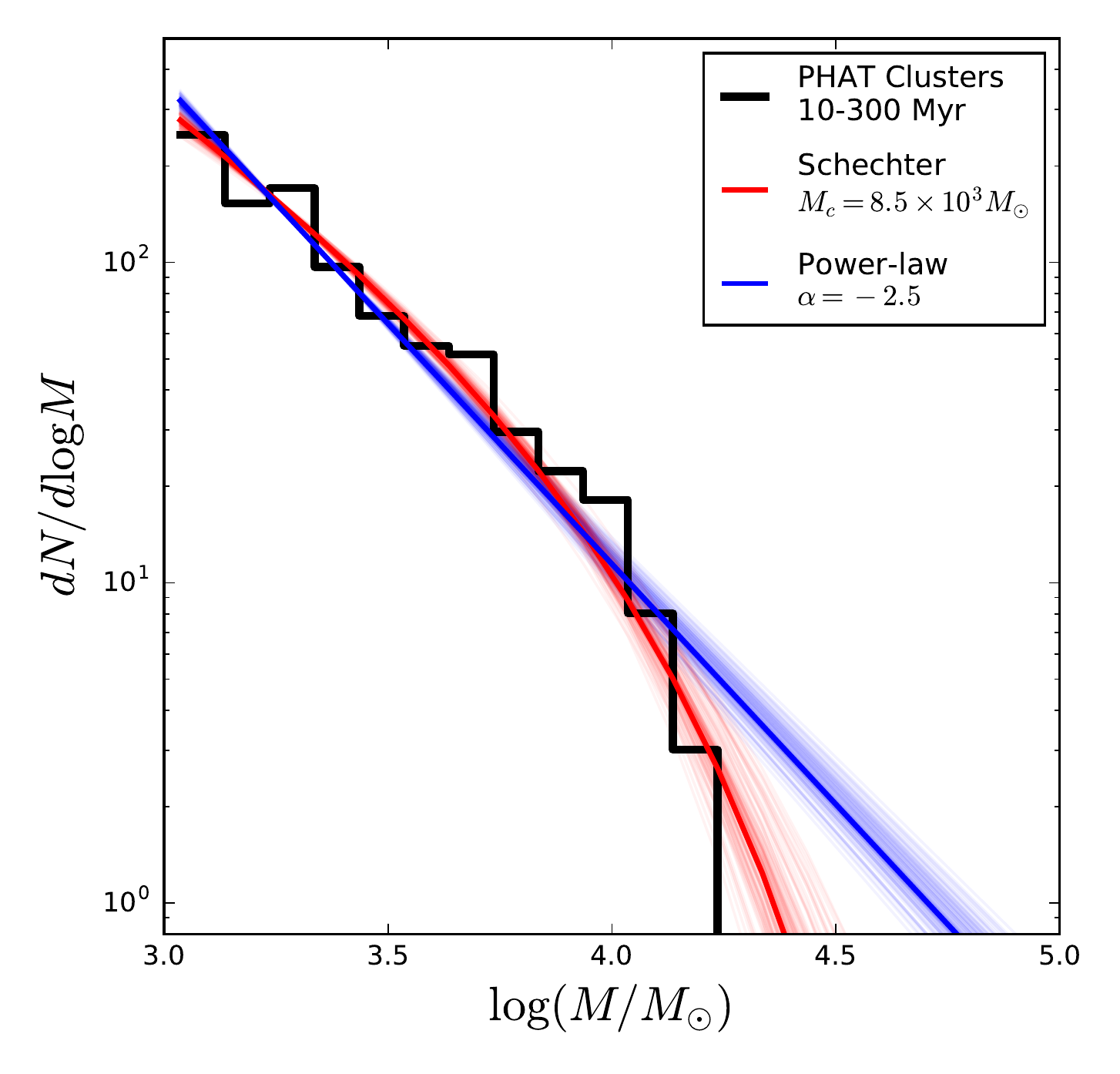}
\includegraphics[width=.34\textwidth]{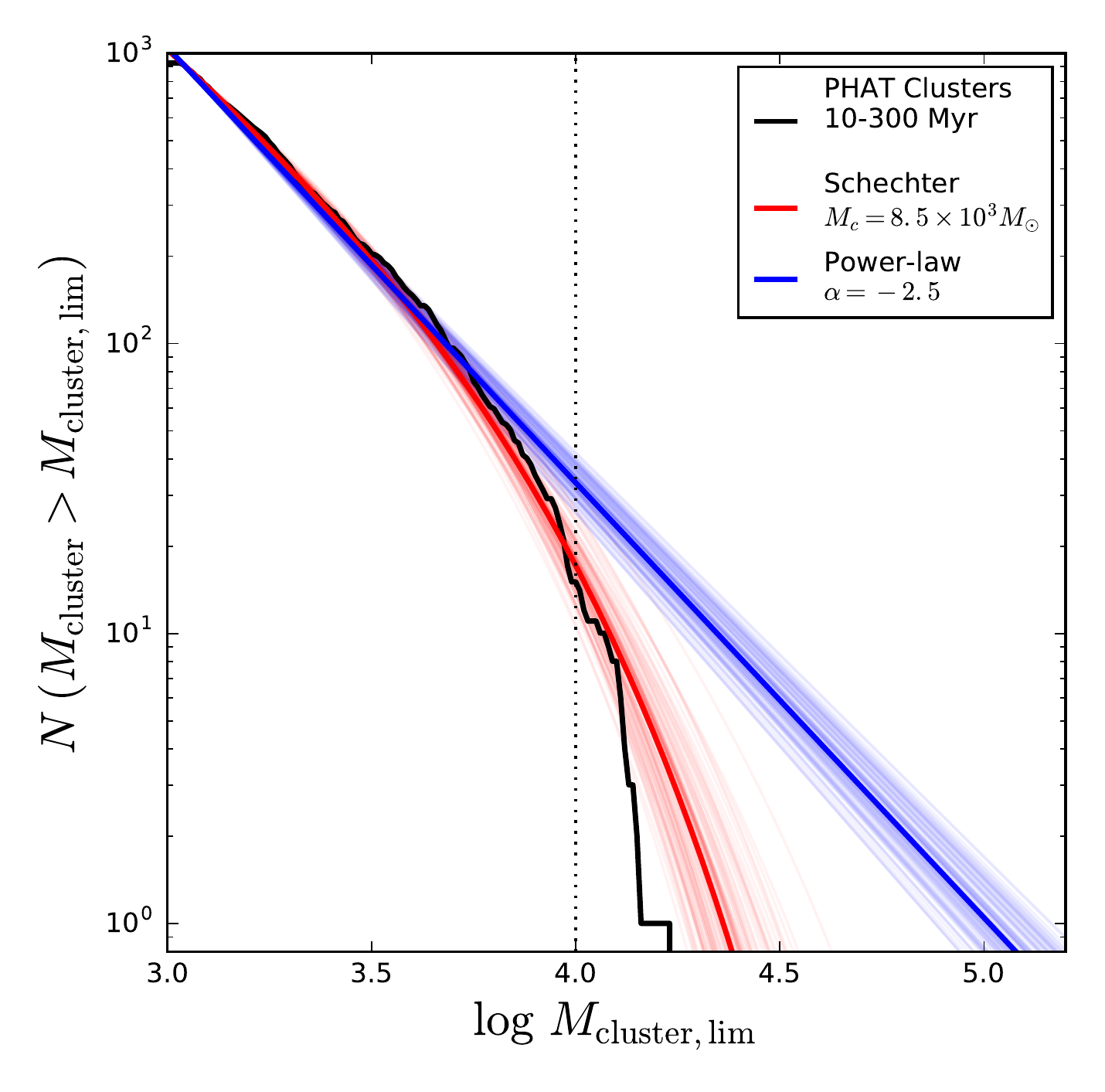}
\includegraphics[width=.3\textwidth]{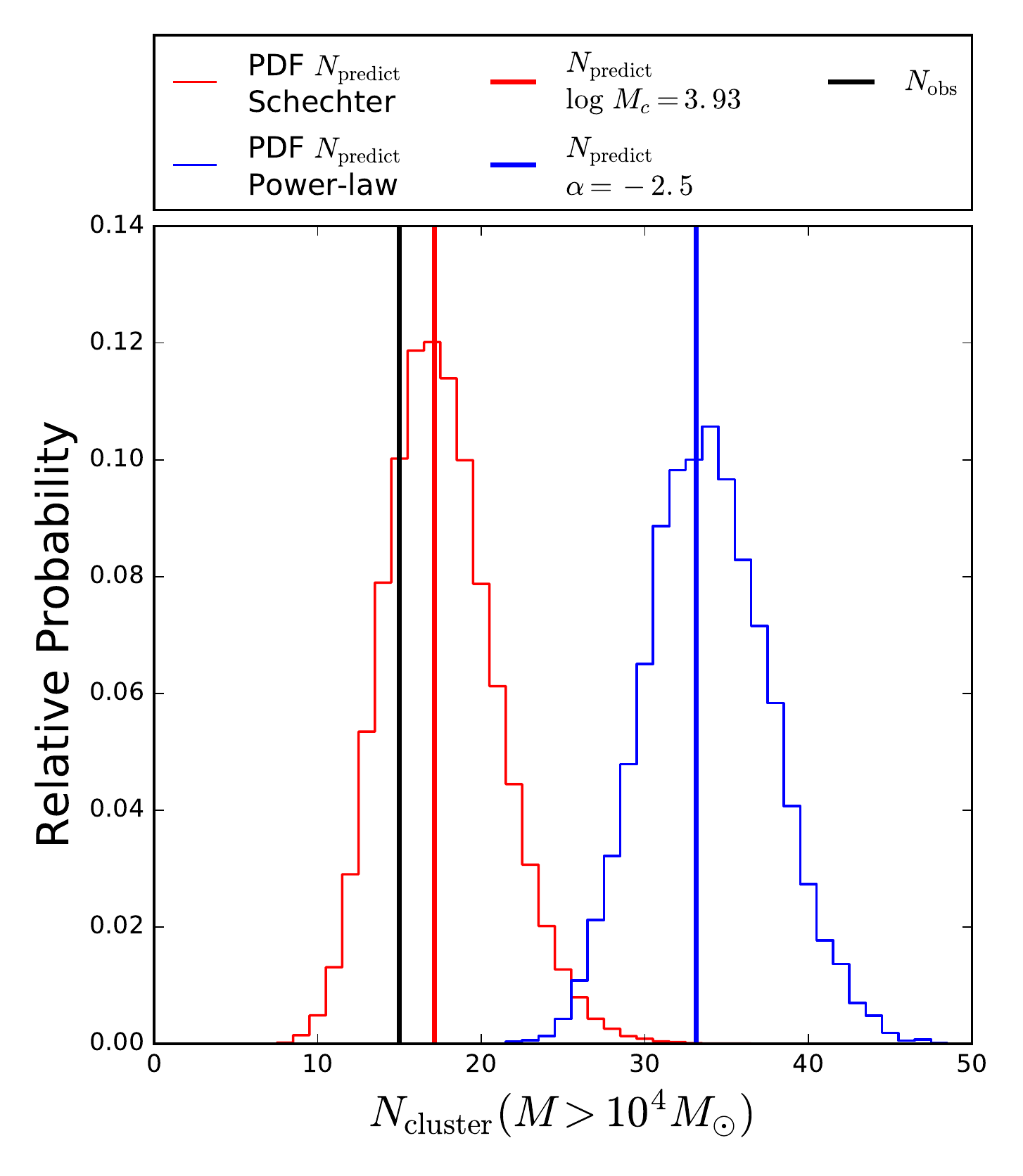}
\caption[Comparison of Functional Forms: Predicted $N_{\mathrm{cluster}}$ ($>$10$^4 M_{\odot})$]{A comparison of Schechter (red) and power-law (blue) mass function fits to the observed, completeness-corrected distribution (black). Left: A comparison of mass function fits to the observed 10-300 Myr cluster mass distribution, where 100 realizations are drawn from the posterior PDFs of each function.  Center: Cumulative curves for mass function models.  The dotted line denotes the 10$^4$ \solmass\ canonical cutoff mass used in model comparison.  Right: PDFs for the number of clusters with $M > 10^4$ \solmass\ based on the normalizations shown in the left panel.  This panel demonstrates that the observed number of massive star clusters in the PHAT sample ($N=15$; black line) is incompatible with the fitted power-law mass function ($N=33^{+5}_{-3}$; blue line), and is better characterized by the fitted Schechter function ($N=17^{+4}_{-3}$; red line).}
\label{fig_fitcomp}
\end{figure*}

The Schechter and power-law functional forms fitted to the observed PHAT cluster mass function yield similar predictions for low-mass clusters, but diverge significantly for high-mass clusters.  We compare the fitting results for the two functional forms in Figure \ref{fig_fitcomp} using differential and cumulative curves in the left and center panels, respectively.  We observe that the fitted power-law function systematically over-predicts the number of massive clusters, whereas the exponential truncation of the Schechter function allows a significantly better fit to the observed distribution.  

As a quantitative illustration of the difference between the fitted Schechter and power-law functions at high cluster mass, we compare the number of clusters with masses $>$10$^4$ \solmass\ observed by PHAT to predictions from the fitted functions.  We transform the posterior PDFs derived for the Schechter and power-law function parameters into PDFs of $N_{\rm cluster}$ with mass greater than $>$10$^4$ \solmass, assuming a normalization set to match the total completeness-corrected number of clusters above a limiting mass of 1080 \solmass.  We plot the resulting PDFs for the Schechter and power-law fits in the right panel of Figure \ref{fig_fitcomp}, and compare these predictions to the observed value of fifteen $>$10$^4$ \solmass\ clusters.

This comparison shows that the 15 observed $>$10$^4$ \solmass\ clusters is incompatible with the $33^{+5}_{-3}$ prediction for the power-law function fit at high significance ($>$4$\sigma$), while well-matched to the $17^{+4}_{-3}$ prediction for the Schechter function.  While this illustration uses an arbitrary threshold cluster mass of 10$^4$ \solmass, we find that the fitted power-law mass function is discrepant at $>$3$\sigma$ significance for any threshold mass greater than 8$\times$10$^3$ \solmass.

We also note that the discrepancy in the observed number of $>$10$^4$ \solmass\ clusters would be even worse for a shallower power-law mass function.  A prediction of 101 $>$10$^4$ \solmass\ clusters, calculated for a canonical $-2$ power-law slope similarly normalized to the total number of clusters above a limiting mass of 1080 \solmass, is clearly discrepant with the observed population of PHAT clusters.

In addition to the specific comparison of Schechter and power law fits at the high mass end, we also compute Kolmogorov-Smirnov (KS) and Anderson-Darling (AD) test statistics and probabilities to assess the overall goodness-of-fit for each functional form to the observed data.  We acknowledge that the Schechter function does not perfectly capture the observed distribution, as shown in the center cumulative distribution panel of Figure \ref{fig_fitcomp}.  A sharper truncation would improve the fit, but the two-parameter Schechter function provides a satisfactory fit to the data.  We derive KS and AD probabilities of 0.376 and 0.430, respectively, demonstrating that the observed data are consistent with our most likely Schechter function.  In contrast, we find KS and AD probabilities for the most likely power-law function of 0.007 and 0.019, respectively, allowing us to discard the null hypothesis that our data were drawn from the most likely power law distribution with high confidence.  Please note that KS and AD probabilities for both functional forms were computed via simulation to properly assess the significance of the test results.

The systematic over-prediction of the massive cluster population by the power-law mass function model argues strongly for the existence of a high-mass truncation of the cluster mass function, and rules out the notion of a universal, pure power-law cluster mass function where the maximum cluster mass is driven only by sampling statistics.  The exponentially-truncated Schechter function serves as a useful description of the high-mass end of the cluster mass distribution, allowing us to compare the M31 mass function to those in other galaxies.

\section{Discussion} \label{discuss}

\subsection{Mass Function Truncations for Young Cluster Systems: Correlation with \SigSFR} \label{discuss_ymc}

In this section we examine whether the physical conditions of star formation in the PHAT survey region of M31 can explain the low value of $M_c$ measured here relative to previous studies.  We combine our M31 mass function result with those from the literature and find a clear correlation between the mass function truncation, $M_c$, and the SFR surface density, \SigSFR.

We complement the PHAT $M_c$ result with young cluster mass function measurements from the literature.  We use the value of $\log (M_c/M_{\odot})$=$6.3^{+0.7}_{-0.3}$ for the Antennae, as calculated by \citet{Jordan07} using the 2.5--6.3 Myr cluster mass distribution data from \citet{Zhang99}. We also use results from \citet{Gieles09} for M51, and survey-wide results from \citet{Adamo15} for M83.  The M51 and M83 measurements are consistent with $M_c \sim 2 \times 10^5$ \solmass, which is the value reported by \citet{Larsen09} for a combined analysis of $\sim$20 nearby spiral galaxies (of which M51 and M83 were members).  We observe that $M_c$ values among the four galaxies vary by $>$2 orders of magnitude.  While the current sample of galaxies with robust Schechter function fits in the literature is relatively small, we benefit greatly from the large dynamic range spanned in characteristic truncation mass and star formation activity.

The four galaxies studied here span a wide range of star formation intensity, from relatively quiescent activity in M31, to merger-induced starburst activity in the Antennae.  M31's low SFR is characteristic of a ``green valley'' galaxy \citep{Mutch11}, and its star formation activity is primarily contained within a 10 kpc star-forming ring, which may be associated with the outer Lindblad resonance of a central bar \citep{Athanassoula06, Blana17}.  The Antennae serve as the prototype for a galaxy merger, providing one of the youngest and closest laboratories for studying high-intensity star formation and massive cluster formation \citep[e.g.,][]{Whitmore10, Johnson15_Antennae}.  In between, M51 and M83 both show signs of recent or on-going galaxy interactions that produce strong present-day star formation, high-amplitude spiral arms, and bar-driven gas flows.

We quantify $M_c$ variations as a function of \SigSFR, an observable metric of star formation intensity.  Unlike an unnormalized galaxy-integrated SFR that scales strongly with global galaxy mass, \SigSFR\ tends to better differentiate galaxies according to differences in local star formation properties.  Furthermore, we calculate galaxy-wide \SigSFR\ values using a SFR-weighted average of kpc-scale \SigSFR\ observations, represented hereafter as \CSigSFR.  This weighted average provides a characteristic, global metric that accurately represents the properties of the local environments in which stars are forming.

We derive new \SigSFR\ measurements for each of the four galaxies in our sample, yielding a homogeneous set of observations that is well-suited for $M_c$-\SigSFR\ correlation analysis.  For each galaxy, we construct a map of \SigSFR\ using a kpc-scale spatial kernel, and obtain a global \CSigSFR\ measurement by computing a SFR-weighted average over the set of local measurements represented in the map.  In addition to the weighted-average, we also report the narrowest percentile range containing 68\% ($\pm1\sigma$) of the SFR-weighted local \SigSFR\ measurements for each galaxy.  This interpercentile range serves as a reminder that the global \CSigSFR\ values we calculate represent an underlying distribution of local star formation environments.

\begin{figure}
\centering
\includegraphics[scale=0.7]{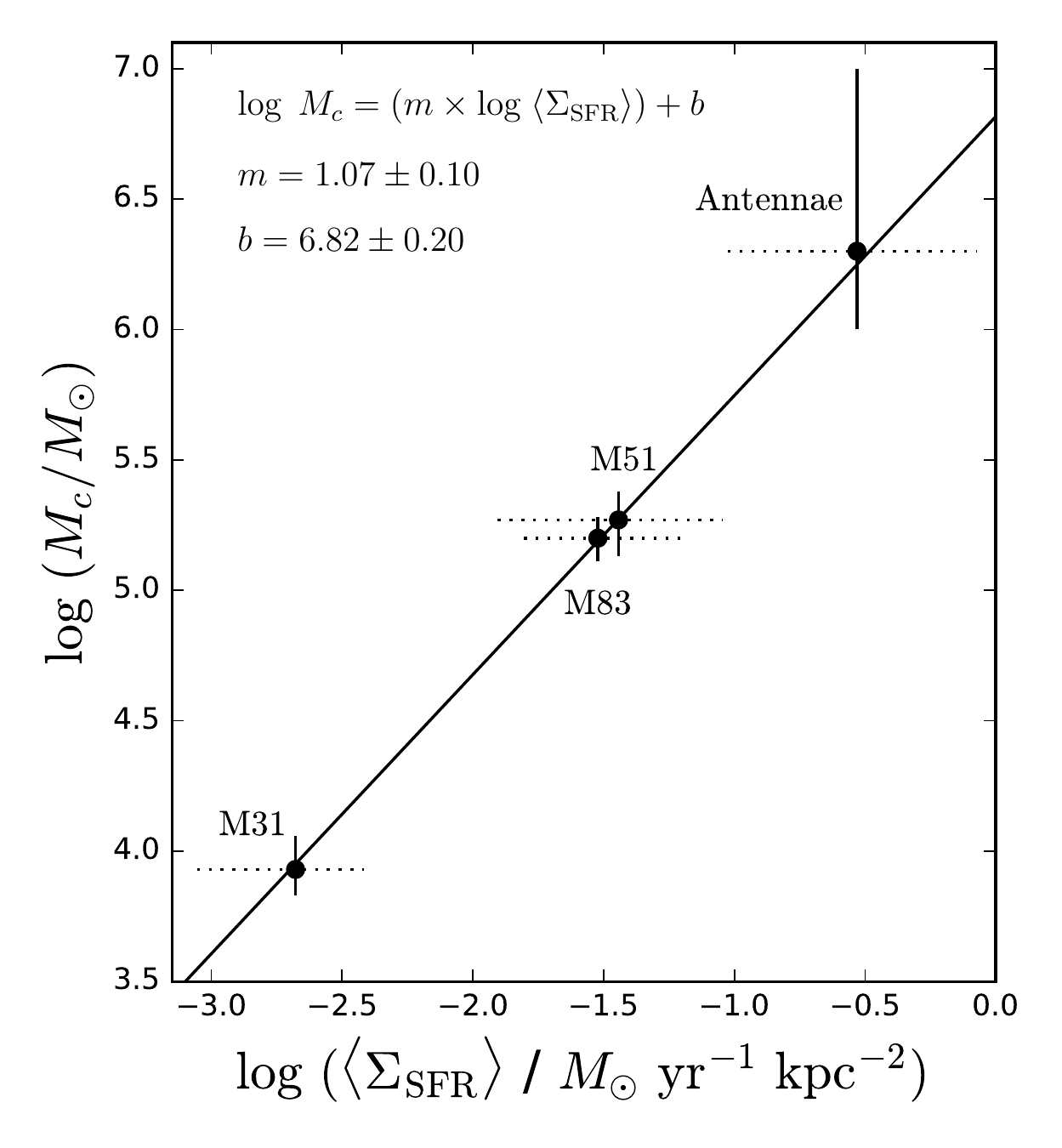}
\caption[$M_c$ Trends: Young Clusters versus \SigSFR]{Comparison of $M_c$ fits for young cluster samples as a function of \SigSFR\ for M31, M83 \citep{Adamo15}, M51 \citep{Gieles09}, and Antennae \citep{Zhang99, Jordan07}.  Solid vertical bars denote fitting uncertainties for $M_c$, and dotted horizontal bars denote the narrowest 68\% interpercentile range of local \SigSFR\ measurements within each galaxy.  We perform a linear fit to the data, and find that $M_c \propto$ \SigSFR$^{1.1}$.}
\label{fig_mc_ymc}
\end{figure}

We present a detailed description of the \SigSFR\ calculations in Appendix \ref{appendix_sigsfr}, including a discussion and justification regarding our use of a SFR-weighted \CSigSFR.  Briefly, we use spatially resolved star formation history maps from \citet{Lewis15} to compute SFR averaged over 10-100 Myr and produce \SigSFR\ maps of the PHAT survey region in M31, following the same methodology used by \citet{Johnson16_gamma}.  For M51, M83, and the Antennae, we use GALEX FUV and Spitzer 24$\mu$m imaging to produce \SigSFR\ maps, following the SFR calibration and methodology used by \citet{Leroy08}.  We present $M_c$ and \CSigSFR\ measurements for the galaxy sample in Table \ref{tbl_mcdata} and plot these results in Figure \ref{fig_mc_ymc}.

\begin{deluxetable*}{lcccc}
\tablecaption{$M_c$ and \SigSFR\ Data \label{tbl_mcdata}}
\tabletypesize{\footnotesize}
\tablehead{\colhead{Galaxy} & \colhead{Region Name} & \colhead{log ($M_c$/\solmass)} & \colhead{log (\CSigSFR/\solperyr\ kpc$^{-2}$)} & \colhead{$M_c$ References}}
\startdata
M31       & PHAT & $3.93^{+0.13}_{-0.10}$  & $-2.68^{+0.26}_{-0.38}$ & This Work \\
M51       & \nodata & $5.27^{+0.11}_{-0.14}$ & $-1.44^{+0.40}_{-0.46}$ & \citealt{Gieles09} \\
M83       & 0.45--4.5 kpc & $5.20^{+0.08}_{-0.09}$ & $-1.52^{+0.34}_{-0.28}$ & \citealt{Adamo15} \\
Antennae  & \nodata & $6.3^{+0.7}_{-0.3}$ & $-0.53^{+0.46}_{-0.49}$ & \citealt{Jordan07} \\
Normal Galaxies  & \nodata & 5.32  $\pm$ 0.10  & \nodata   & \citealt{Larsen09}
\enddata
\tablecomments{\CSigSFR\ results reflect a SFR-weighted galaxy-wide average of local measurements observed at 2--3 kpc$^2$ scale, while upper and lower limits bracket the narrowest 68\% interpercentile range of local \SigSFR\ measurements.}
\end{deluxetable*}

Figure \ref{fig_mc_ymc} shows a strong correlation between $M_c$ and \CSigSFR, spanning $>$2 orders of magnitude in each quantity.  The observed trend suggests a strong dependence of the cluster mass function truncation on the characteristics of the galactic star forming environment. We quantify the observed relationship between $M_c$ and \CSigSFR\ by fitting a linear relation to the observed data in log $M_c$--log \CSigSFR\ parameter space:
\begin{equation} \label{eq_mc_sigsfr}
\log M_c = (1.07 \pm 0.10) \times \log \langle\Sigma_{\mathrm{SFR}}\rangle + (6.82 \pm 0.20).
\end{equation}
The fitting suggests a near-linear proportionality between the mass function truncation and \CSigSFR, such that $M_c \propto$ \CSigSFR$^{1.1}$.  The quoted uncertainties on the fitted slope account for $M_c$ uncertainties only; uncertainties on the slope  increase to $\pm0.2$ if 0.2 dex uncertainties on \CSigSFR\ measurements were included, or they would increase to $\pm0.3$ if the 68\% interpercentile range is used to define the \SigSFR\ confidence interval.

The $M_c$-\SigSFR\ relation we identify here is defined at galaxy-integrated scales.  This choice of averaging scale provides the large star cluster number statistics required to place strong constraints on cluster mass function shape and the presence of a high-mass truncation.  However, galaxy-wide averaging obscures the complexity of physical dependencies related to star cluster formation, beyond the \SigSFR\ dependence we characterize here.  Therefore, we stress caution when extrapolating cluster formation behavior at smaller scales using Equation \ref{eq_mc_sigsfr}.  Spatially-resolved observational follow-up work examining the physical drivers of massive cluster formation has the potential to further our understanding of star formation in high gas density, high star formation efficiency environments, but obtaining statistical robust constraints in the regime of small cluster number statistics will be a significant challenge (as discussed in Section \ref{results_radial}).

We note that $M_c$ and cluster formation efficiency ($\Gamma$ = $M_{\rm cluster}$/$M_{\rm total}$) have both been shown to vary systematically with \SigSFR, and seek to clarify the interconnected yet distinct nature of these variations.  Assuming a fixed normalization for the low-mass end of a Schechter mass function, decreasing $M_c$ tends to decrease the integrated stellar mass of a cluster population.  As a result, the observed variation in $M_c$ leads to correlated declines in both $\Gamma$ and \SigSFR.  However, only a small fraction of the total $\Gamma$ variation observed can be explained by the variation in $M_c$ alone.  For example, decreasing $M_c$ from $10^6$ to $10^4$ \solmass\ only produces a factor of $\sim$2 change in $\Gamma$, while observations and theoretical predictions show evidence for more than an order of magnitude change over the same range of \SigSFR\ \citep{Kruijssen12, Johnson16_gamma}.  Therefore, variations in $\Gamma$ do not stem solely from differences in high mass cluster formation as a function of \SigSFR, but reflect broad differences in cluster formation over a wide range of masses.

\subsubsection{Physical Drivers of $M_c$-\SigSFR\ Correlation: Pressure} \label{discuss_pmp}

We explore the role that interstellar pressure may play in driving the observed $M_c$-\SigSFR\ correlation.  Large stellar densities observed in massive clusters and globular clusters suggest extremely high gas densities in progenitor molecular clouds at the time of formation \citep{Elmegreen97}.  Maintaining such high densities is likely to require large external pressures to keep the natal gas confined, which motivates our specific interest in pressure over other environmental parameters.  While the coupling between external and internal pressures for host molecular clouds is currently debated, observational evidence favoring the influence of galactic environment and external pressure on molecular cloud properties has begun to emerge \citep{Hughes13, Colombo14}. These confining pressures may be set by the equilibrium conditions of star-forming disks, or may be produced transiently over large spatial scales in galaxy mergers \citep{Renaud15_Antennae} or over small scales in molecular cloud collisions \citep{Fukui14}.

For the simple case of an equilibrium star-forming disk, we can examine whether observed variations in $M_c$ are consistent with the predicted scaling behavior of pressure as a function of \SigSFR. We approximate the dependence between mid-plane pressure ($P_{\rm{mp}}$) and \SigSFR\ for the case of a stable star-forming galaxy disk following the logic presented in \citet{Elmegreen09}.  We combine the expectation that $P_{\rm{mp}}$ scales as \SigGas$^2$ with the empirical Kennicutt-Schmidt relation \citep{Kennicutt98} where \SigSFR\ $\propto$ \SigGas$^{\rm{1.4}}$, and we predict that $P_{\rm{mp}} \propto$ \SigSFR$^{\rm{1.4}}$.  This predicted dependence is steeper than the observed trend, where $M_c$ $\propto$ \SigSFR$^{1.1}$, suggesting that transient enhancements of interstellar pressure or other environmental characteristics drive the behavior of high-mass cluster formation.

Establishing that pressure, or another physical driver, is responsible for the mass function truncation variations will require additional study.  Rather than relying on indirect scaling arguments, obtaining observational estimates of interstellar pressure and other environmental variables and directly analyzing their correlation with $M_c$ observations could help identify key galactic properties that influence massive cluster formation behavior.

Another avenue of study involves the comparison of the star cluster mass function with the giant molecular cloud (GMC) mass function.  As clusters are formed out of molecular gas, and the GMC mass function is known to vary with galactic environment \citep[e.g.,][]{Colombo14}, understanding the connection between the behavior of these two mass distributions may shed light on the underlying physics involved.  To this point, \citet{Kruijssen14} suggested that the maximum mass scale of both star clusters and GMCs might have a common origin, tied to the Toomre mass \citep{Toomre64}.   The CARMA survey of M31 GMCs (A. Schruba et al., in preparation) and other extragalactic GMC surveys with ALMA and other facilities will provide many opportunities to study the connection between cluster and molecular cloud mass functions in detail, and to test theoretical explanations for observed behavior.

\subsection{Mass Function Truncations for Globular Cluster Systems: Similarity to Young Clusters?} \label{discuss_gc}

Old globular cluster systems have a dramatically different mass function shape compared to the young cluster systems discussed in the previous section.  The globular cluster mass function (GCMF) is commonly parameterized using a Gaussian or log-normal form, and shows a clear peak at a near-constant mass of $2\times10^5$ \solmass\ \citep[e.g.,][]{Jordan07, Villegas10}.

Early theoretical work proposed that globular clusters formed in a way that differs from young clusters forming today, following a mass distribution which peaks at a characteristic mass scale \citep[e.g.,][]{Peebles68, Fall85}.  In contrast, more recent work has argued that globular clusters form with an initial power-law (or Schechter function) mass distribution that evolves to a peaked distribution due to dynamical evolution and destruction processes \citep[e.g.,][]{Gnedin97, Fall01, Kruijssen15}.  The use of an initial power-law mass function is motivated by cluster formation behavior observed at low redshift, and assumes cluster formation proceeds similarly at all redshifts.  In this case, globular cluster populations today are the surviving relics of a population that formed in the same way that young massive clusters do in the present day.  The small number of young massive clusters presently formed at low redshift, relative to the large number of old massive globular clusters, results from an overall decline of the cosmic star formation history since $z\sim2$ \citep{Madau14}, leading to a corresponding decline in local \SigSFR\ and massive cluster formation.

Globular cluster systems in early-type galaxies show systematic variations in their luminosity function shapes.  The width of the peaked luminosity functions are observed to increase with host galaxy mass, as observed for Virgo cluster members \citep{Jordan06, Jordan07}, Fornax cluster members \citep{Villegas10}, and seven brightest cluster galaxies in other massive galaxy clusters \citep{Harris14}.  \citet{Jordan07} demonstrate that this increase in width of the globular cluster luminosity function, and subsequently the GCMF, can be interpreted either as an increase in the dispersion ($\sigma_{\rm LN}$) of a traditional log-normal functional form, or as an increase in $M_c$ for an evolved Schechter function --- a functional form inspired by \citet{Fall01} that accounts for cluster mass loss.  This behavior is broadly similar to the mass function variations observed for young cluster systems.  We therefore compare these two sets of $M_c$ measurements and investigate a possible connection between globular cluster and young cluster formation.  If the two cluster populations follow the same $M_c$--\SigSFR\ correlation, this could signal they form through a common formation pathway.

\begin{figure*}
\centering
\includegraphics[scale=0.55]{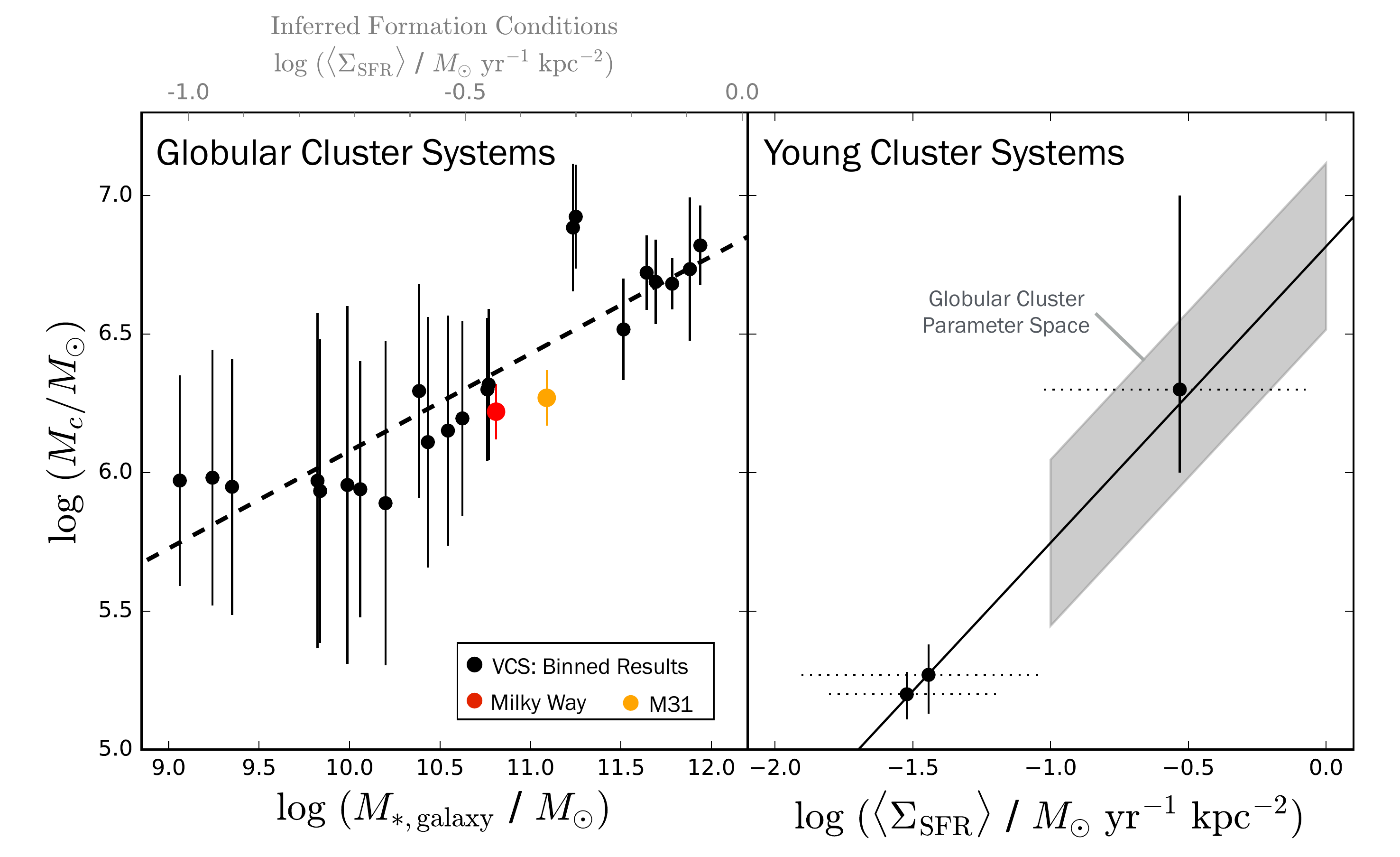}
\caption[$M_c$ Trends: Globular Clusters and Young Clusters]{Left: Comparison of $M_c$ fits for old globular cluster systems as a function total present-day stellar mass of the host galaxy.  We plot the binned results from the ACS Virgo Cluster Survey (VCS) as well as the Milky Way result presented in \citet{Jordan07}.  Right: $M_c$ results for young cluster samples as a function of \CSigSFR, as presented in Figure \ref{fig_mc_ymc}.  We highlight a portion of the $M_c$--\CSigSFR\ relation that corresponds to the range of $M_c$ values observed for globular cluster systems. Correspondingly, we overlay a second \CSigSFR\ axis along the top of the left panel that shows the inferred \CSigSFR\ of globular cluster formation environments, assigned by combining the correlations from the left and right panels.}
\label{fig_mc_gc}
\end{figure*}

\subsubsection{Globular Cluster $M_c$ Measurements} \label{discuss_gc_data}

We compare the young cluster $M_c$ measurements from Section \ref{discuss_ymc} to globular cluster $M_c$ measurements from the ACS Virgo Cluster Survey \citep[VCS;][]{Cote04} published by \citet{Jordan07}. These authors fit the data using an evolved Schechter function, allowing a direct comparison between the two sets of results\footnote{Results from \citet{Villegas10} and \citet{Harris14} are not included because only Gaussian function fitting results are published.  While these additional fits would boost the sample's number statistics, the \citet{Jordan07} results are representative of the larger, combined dataset.}.

We plot globular cluster $M_c$ values as a function of present-day host galaxy mass in the left panel of Figure \ref{fig_mc_gc}, reproducing the data and result from \citet{Jordan07}.  These data points reflect binned results based on $z$-band luminosity function fitting, where globular cluster systems for small subsets of galaxies (1--9; see their Table 3) are stacked to boost cluster number statistics.  We obtain $M_c$ masses by converting luminosity function fitting results into cluster mass parameter space via $z$-band mass-to-light ratios ($\Upsilon_z$) derived from SSP models, then transform from present-day to initial stellar mass by accounting for stellar evolution-based mass loss and death.  We use the Flexible Stellar Population Synthesis code \citep[FSPS;][]{Conroy09, Conroy10} to calculate the conversions, where $\Upsilon_z$ are based on a nominal 13 Gyr cluster age and a \citet{Kroupa01} stellar IMF, and are relatively insensitive to metallicity in agreement with \citet{Jordan07}.  The conversion from present-day to initial cluster mass accounts for the $\sim$45\% of SSP mass returned to the ISM over the nominal 13 Gyr cluster lifetime.  We derive galaxy stellar masses ($M_{*\rm{, galaxy}}$) using $z$-band luminosities from \citet{Ferrarese06}, ($g-z$) colors and distances from \citet{Blakeslee09}, and color-based stellar mass-to-light ratios from \citet{Into13}.

We fit the following linear relation for $M_c$ as a function of $M_{*\rm{, galaxy}}$:
\begin{multline} \label{eq_mc_mstar}
\log (M_c/M_{\sun}) = (0.35 \pm 0.07) \times \log (M_{* \mathrm{, galaxy}}/M_{\sun}) \\
+ (2.6 \pm 0.8).
\end{multline}
We emphasize that the observed correlation is unlikely to be directly linked to stellar mass.  Instead, we expect that the stronger underlying correlation is that the intensity of star formation is higher in progenitor galaxies that go on to form more massive galaxies as compared to progenitor galaxies that merge to form less massive galaxies.  In this scenario, present day host galaxy mass serves as a proxy for galaxy assembly history.

We also compare mass function constraints for the Milky Way and M31 globular cluster systems to demonstrate that the GCMF behavior shown here is not a special feature of early-type host galaxies in galaxy cluster environments.  We use the Milky Way $M_c$ reported in \citet{Jordan07}, corrected using the $\Upsilon_z$ values derived above for the VCS measurements, and pair it with the Galactic stellar mass determination from \citet{McMillan11}.  We performed our own globular cluster luminosity function fit for M31, as described in Appendix \ref{appendix_m31gcmf}, and pair this $M_c$ measurement with the M31 stellar mass from \citet{Tamm12}.  These data points are not included in the $M_c$--$M_{*\rm{, galaxy}}$ fit, but appear to follow a similar trend as found for the VCS galaxies.  The deviation of the late-type spiral galaxies toward higher $M_{*\rm{, galaxy}}$ with respect to the relation for early-type hosts is plausibly explained by differences in galaxy evolution.  Progenitor galaxies with similar properties, and $M_c$ values, at the epoch of globular cluster formation will diverge in terms of present day stellar mass if their star formation histories differ significantly, as explained by \citet{Mistani16} for the case of field versus galaxy cluster dwarf galaxies.

\subsubsection{$M_c$ Comparison and \CSigSFR\ Predictions} \label{discuss_gc_results}

The VCS globular cluster systems span a $\sim$1 dex range in $M_c$, from $\sim$10$^{6}$--10$^{7}$ \solmass, which overlaps with the upper range of $M_c$ values observed in young cluster systems (right panel of Figure \ref{fig_mc_gc}).  Given the comparable $M_c$ values and existing models which assume young massive cluster formation and globular cluster formation are governed by the same physical processes, we hypothesize that the same $M_c$--\SigSFR\ relation observed for young cluster systems also holds for globular cluster systems.  If true, then globular clusters in these early-type galaxies must have formed in star forming environments with \CSigSFR\ values between 0.1--1.0 \solperyr\ kpc$^{-2}$ --- within $\pm$0.5 dex of the Antennae's \CSigSFR.  We highlight this portion of the $M_c$--\SigSFR\ correlation in the right panel of Figure \ref{fig_mc_gc} with a gray box.  Furthermore, we combine the $M_c$--$M_{*\rm{, galaxy}}$ and $M_c$--\SigSFR\ correlations and infer properties of globular cluster formation environments as a function of present day host galaxy mass, presented as a upper x-axis in the left panel of Figure \ref{fig_mc_gc}.

The $M_c$ values ascribed to the globular cluster systems, and hence the \CSigSFR\ values assigned according to the young cluster $M_c$--\SigSFR\ relation presented in Section \ref{discuss_ymc}, depend on assumptions made about globular cluster mass loss.  We explicitly account for stellar evolution-based mass loss in this study through $\Upsilon_z$, and the constant mass loss term, $\Delta$, is included as part of the evolved Schechter function parameterization\footnote{Please note that the fitted $M_c$ values in the evolved Schechter function parameterization represent initial values and do not need to be corrected for $\Delta$-parameterized mass loss.} to account for additional sources of mass loss.  The \citet{Jordan07} fitting results for $\Delta$ call for negligible mass function evolution at the high-mass end, in agreement with predictions for most forms of globular cluster mass loss.  However, globular cluster formation models developed to explain multiple population phenomena call for large amounts of mass loss ($>$90\% of initial mass) in order to explain the observed ratio of enriched to unenriched populations and obtain the necessary dilution of enriching material \citep[e.g.,][]{DErcole08, Conroy12}.  Inferred $M_c$ and \CSigSFR\ values would increase in the case of large, cluster mass-independent mass loss.  Recent observational studies disfavor formation models with large fractional mass loss \citep{Larsen12, Larsen14, Bastian15, Schiavon17_Nrich}, but the matter is far from settled.

The 0.1--1.0 \solperyr\ kpc$^{-2}$ \CSigSFR\ values for globular cluster formation inferred here are lower than the most extreme values observed in intense starbursts and luminous infrared galaxies: 10--100 \solperyr\ kpc$^{-2}$ \citep[e.g.,][]{Kennicutt12}.  The lower \CSigSFR\ predictions are partially explained by their galaxy-averaged spatial scales, where the underlying distribution includes more extreme \SigSFR\ values in smaller, localized regions.  However, the predicted 0.1--1.0 \solperyr\ kpc$^{-2}$ range is not unreasonable when considering that a significant fraction of globular clusters (especially metal-poor systems) form in lower-mass progenitors before merging and accreting onto more massive halos, and these progenitors are not expected to host 10--100 \solperyr\ kpc$^{-2}$ starburst conditions.

To place these \CSigSFR\ predictions into context, we highlight a number of numerical simulation studies that make related predictions about the properties of globular cluster formation.  In a study by \citet{Peng08} investigating the specific frequency ($S_N$) of relatively low-mass ($M_z > -19$; $\log (M_{*}/M_{\sun}) < 9.6$) Virgo cluster galaxies, the authors examined theoretical predictions for star and cluster formation histories from the Millennium simulation \citep{Springel05}.  We select a comparable sample of low-mass ($\log (M_{*}/M_{\sun}) < 9.6$) VCS galaxies and find an average truncation mass of log ($M_c$/\solmass) $\sim 5.9$.  Paired with simulation-based predictions of log (\SigSFR/\solperyr\ kpc$^{-2}$) $\sim -1.5$ at the $z\sim4.5$ peak globular cluster formation epoch, the resulting prediction lies to the left of the observed $M_c$--\SigSFR\ relation.  However, this result relies on a large number of assumptions (e.g., semi-analytic star formation prescriptions, approximate galaxy size estimates) that may bias the \SigSFR\ prediction.  The latest generation of cosmological simulations that include full baryonic physics \citep[e.g., the Illustris simulation;][]{Vogelsberger14} and high-resolution zoom-in galaxy simulations \citep[e.g., the FIRE simulations;][]{Hopkins14, Hopkins17} motivate a new look at \SigSFR\ predictions during the epoch of globular cluster formation, building on studies that already explore kpc-scale \SigSFR\ at high redshift \citep{Orr17} and the impact of galaxy environment on globular cluster formation \citep{Mistani16}.

Another recent numerical simulation study by \citet{Li17} is also closely related to our exploration of globular cluster formation, $M_c$ values, and influence of star-forming environment.  The authors implement star cluster-based star formation in a cosmological  galaxy formation simulation and find that the resulting cluster initial mass function is well-described by a Schechter function.  \citet{Li17} find a positive correlation between the characteristic truncation mass and SFR for high redshift ($z > 3$) cluster formation, such that $M_c \propto \rm{SFR}^{1.6}$.  While we encourage future comparisons based on \SigSFR\ rather than an unnormalized SFR (see discussion in Section \ref{discuss_ymc}), we find a similar correlation between $M_c$ and integrated SFR for our four galaxy sample, such that $M_c \propto \rm{SFR}^{1.5}$.  This result supports our hypothesis that globular cluster formation at high redshift follows similar relations as young cluster formation at the present day.  Further examination of the cluster mass function behavior in the context of theoretical globular cluster formation models \citep[e.g.,][]{Kruijssen15,Renaud17} is clearly desirable.

A universal correlation between cluster formation and star formation environment has important implications.  If this hypothesis is true, measurements of the upper end of the GCMF could allow observers to infer important details about the hierarchical build-up of galaxies and the physical conditions of star formation in the early universe through studies of globular cluster systems.  There are many aspects of the globular cluster formation we are yet to fully understand (e.g., their specific frequencies, metallicity distributions, destruction and mass loss mechanisms, the origin of He and light-element abundance variations within individual clusters), but $M_c$ measurements could serve as an important tool for studying star formation in the early Universe.

\section{Summary} \label{summary}

We find evidence for a high-mass truncation of the star cluster mass function within the PHAT survey region in M31.  Parameterized using a Schechter function, this exponential truncation has a characteristic mass of $M_c$ = $8.5^{+2.8}_{-1.8} \times 10^3$ \solmass.  This truncation mass is the lowest value ever observed for a star cluster population, and provides strong evidence of an upper mass limit for the PHAT cluster sample that rules out a universal power-law cluster mass distribution where the maximum cluster mass is set by sampling statistics.

When we combine the M31 mass function fit and previous $M_c$ results for young cluster systems from the literature, we identify a strong systematic correlation between the truncation mass of the star cluster mass function and star formation environment, as characterized by \SigSFR.  The characteristic truncation mass increases with increasing star formation intensity, such that $M_c \propto$ \CSigSFR$^{\sim1.1}$.  This scaling relation might suggest an underlying physical dependence driven by interstellar pressure, but further study is required to confirm the relationship and its physical underpinnings.

Finally, we highlight that globular cluster systems also show systematic variations in the high-mass truncation of their mass distributions.  We hypothesize that these mass function variations are the result of the same environmentally-dependent truncation relation that we observe for young cluster systems in nearby galaxies.  This proposed commonality between ancient globular clusters and young clusters forming today could represent a long-sought link demonstrating that, while star formation in the early Universe was generally more active and intense, star cluster formation follows the same universal trends across all of cosmic time.  Furthermore, it could enable the use of a galaxyÕs globular clusters systems to make quantitative statements about its early formation environment.


\acknowledgements
{We recognize and thank the $\sim$30,000 Andromeda Project volunteers who made this work possible.  Their contributions are acknowledged individually at \url{http:// www.andromedaproject.org/\#!/authors}.  We thank Nelson Caldwell, Dimitrios Gouliermis, Raja Guhathakurta, Diederik Kruijssen, S{\o}ren Larsen, Andreas Schruba, and Evan Skillman for their comments and feedback on the paper.  We thank the referee for a useful and helpful report.  Support for this work was provided by NASA through grant number HST-GO-12055 from the Space Telescope Science Institute, which is operated by AURA, Inc., under NASA contract NAS5-26555.  GALEX data presented in this paper were obtained from the Mikulski Archive for Space Telescopes (MAST). Support for MAST for non-HST data is provided by the NASA Office of Space Science via grant NNX09AF08G and by other grants and contracts.  This work also makes use of the NASA/IPAC Infrared Science Archive, which is operated by the Jet Propulsion Laboratory, California Institute of Technology, under contract with NASA.  This work made use of the \texttt{python-fsps} software package \citep{ForemanMackey14_FSPS}.}


\appendix

\section{Calculating a Characteristic Galaxy-averaged \SigSFR} \label{appendix_sigsfr}

\subsection{Motivation}

Defining a robust, galaxy-averaged \SigSFR\ is important when exploring the link between cluster mass function truncation measurements and \SigSFR\ on galaxy-wide scales.  \SigSFR\ is known to vary by more than an order of magnitude within galaxies as measured on 0.1--1.0 kpc scales \citep[e.g.,][]{Leroy08}.  Galaxy-wide \SigSFR\ measurements are often calculated simply by dividing a global SFR by an estimate of total galaxy area \citep[e.g., using $R_{25}$;][]{Larsen00}.  These estimates make an implicit assumption that star formation is distributed uniformly across the galaxy, leading to \SigSFR\ estimates that are biased toward small values due to the centrally-concentrated and clumpy spatial distribution of star formation within galaxies.

In this work, we calculate SFR-weighted \SigSFR\ values, denoted here as \CSigSFR, to provide characteristic, galaxy-integrated measurements that are useful for comparing star formation behavior across our sample of galaxies.  This weighted average accurately summarizes the distribution of local, kpc-scale properties of star formation within a galaxy, accounting for the fact that a large fraction of stellar mass forms in a small fraction of the galaxy area --- in regions that lie in the upper tail of the \SigSFR\ distribution.  We highlight that this averaging technique is conceptually similar to the analysis techniques employed by \citet{Leroy16} in their analysis of gas surface density and ISM properties.

The \CSigSFR\ measurement is defined by two scales: a 2--3 kpc$^2$ measurement scale, and a full-galaxy averaging scale.  The choice of a uniform, sample-wide measurement scale minimizes scale-dependent differences in \CSigSFR\ across the sample, and is set by the available spatial resolution of the SFR tracer observations for the sample's distant galaxies.  In addition, our adoption of a $\gtrsim$1 kpc measurement scale minimizes potential biases on SFR estimates due to the effects of discreteness and non-constant star formation histories when sampling smaller spatial scales and integrated SFRs \citep[see e.g.,][]{Schruba10}.  The choice of galaxy-integrated averaging is driven by the need for large samples of star clusters in order to obtain statistically-significant constraints on the shape of the cluster mass function at its high mass end.

The \CSigSFR\ values we derive represent an underlying distribution of local \SigSFR\ values and star-forming conditions.  This fact becomes particularly important when considering how galaxy-scale correlations based on galaxy-averaged kpc-scale \CSigSFR\ presented in this paper translate to cloud-scale cluster formation behavior.  We stress caution when extrapolating behavior across dissimilar spatial scales.

\subsection{\SigSFR\ Calculations}

We characterize local \SigSFR\ distributions and global average values following steps described in Section \ref{discuss_ymc}.  We begin by integrating SFRs over a common spatial scale (2--3 kpc$^2$) within each of the four galaxies in our sample to map \SigSFR\ locally.  Next, we derive a SFR-weighted average and accompanying interpercentile range to represent the distribution of local \SigSFR\ measurements.  We use the narrowest percentile range that contains 68\% of the weighted \SigSFR\ measurements to characterize the dispersion of the distribution.

Galaxy-specific observational data and SFR estimation techniques fall into two groups: nearby galaxies (M31) and distant galaxies (M83, M51, Antennae).  For M31, we measure \SigSFR\ using the same procedure used in \citet{Johnson16_gamma}: maps of \SigSFR\ were created by averaging the 10-100 Myr star formation history derived from CMD fitting \citep{Lewis15} using a deprojected circular 2 kpc$^2$ tophat spatial kernel.  The use of a larger spatial kernel in this work versus the results published in \citet{Johnson16_gamma} has little effect on the derived distribution nor the weighted average of \SigSFR.  The use of a 2 kpc$^2$ as opposed to 0.5 kpc$^2$ kernel ($r \sim 0.8$ kpc versus 0.4 kpc) results in a 0.05 dex reduction in characteristic \SigSFR.

For M83, M51, and the Antennae, we process GALEX and Spitzer imaging to produce \SigSFR\ maps based on the FUV+24$\mu$m SFR calibration from \citet{Leroy08}.  The images were downloaded from the IRSA and MAST data archives, and include data products produced by the Local Volume Legacy survey \citep{Dale09} and MIPS Local Galaxy Survey \citep{Bendo12}.  We convolve the FUV and 24$\mu$m images to a common 11 arcsec resolution and combine their flux densities according to Equation D11 from \citet{Leroy08} to produce \SigSFR\ maps.  We use a common 3 kpc$^2$ tophat spatial kernel ($r \sim 1$ kpc) to compute matched-resolution local \SigSFR\ measurements and a SFR-weighted average \SigSFR\ for each of the three galaxies.  We note that this newly derived \CSigSFR\ measurements differ from previous values presented in the literature.  These differences occur for a variety of reasons, primarily driven by our use of SFR-weighted averaging, and differences in choice of SFR tracer and calibration.

\section{$M_c$ for M31 Globular Cluster System} \label{appendix_m31gcmf}

We performed an evolved Schechter function fit to the M31 globular cluster system to provide a $M_c$ measurement that can be compared to the VCS results presented by \citet{Jordan07}.  The evolved Schechter function takes the form
\begin{equation}
p_{\textrm{MF}}(M|M_c, \Delta) \propto \frac{1}{(M+\Delta)^2}\ \exp \left ( - \frac{M+\Delta}{M_c} \right ),
\end{equation}
where $M$ is cluster mass, $M_c$ is the characteristic truncation mass, and $\Delta$ is the cumulative mass loss.
We fit the luminosity-based version of this function
\begin{equation}
p_{\textrm{MF}}(m|m_c, \delta) \propto \frac{10^{-0.4(m-m_c)}}{(10^{-0.4(m-m_c)}+10^{-0.4(\delta-m_c)})^2}\ \exp ( -10^{-0.4(m-m_c)} ),
\end{equation}
where $m \equiv C - 2.5 \log M$, $\delta \equiv C - 2.5 \log \Delta$, and $m_c \equiv C - 2.5 \log M_c$ represent magnitude-based versions of the mass-based variables, and C is the conversion factor defined by the solar absolute magnitude and the adopted cluster mass-to-light ratio.

We use the photometry catalog of old globular clusters from \citet{Peacock10} and fit the $z$-band luminosity function using the same probabilistic framework described in Section \ref{analysis}.  We assume $\Upsilon_z$ of 3.0 $M_{\sun}$/$L_{\sun}$ to transform our fitting results back into mass space under the same assumptions used for the VCS $M_c$ data points (i.e., calculated using FSPS with 13 Gyr nominal age, \citealt{Kroupa01} IMF, and correction for stellar evolution mass losses to initial mass).  As a result, we find
\begin{equation}
\log (M_c / M_{\sun}) = 6.27^{+0.11}_{-0.10};\ \log (\Delta / M_{\sun}) = 6.05^{+0.14}_{-0.11} .
\end{equation}
We pair this $M_c$ measurement with a M31 stellar mass determination from \citet{Tamm12} of $\log(M_*/M_{\sun})=11.1 \pm 0.09$ and use these observations to place M31 on the left panel of Figure \ref{fig_mc_gc}.


\bibliographystyle{aasjournal}
\bibliography{}


\end{document}